\documentclass[aps,prd,nofootinbib,twocolumn,showpacs]{revtex4-1}

\pdfoutput=1
\usepackage{graphicx}
\usepackage{color}
\usepackage{textcase}
\usepackage{amsmath, amssymb, amsthm, amsfonts}
\usepackage{enumerate}
\usepackage{relsize} 
\usepackage{array} 
\usepackage{bbm} 
\usepackage{hyperref} 

\def\mbf#1{\ensuremath{\mathchoice{\mbox{\boldmath$\displaystyle#1$}}
{\mbox{\boldmath$\textstyle#1$}}
{\mbox{\boldmath$\scriptstyle#1$}}
{\mbox{\boldmath$\scriptscriptstyle#1$}}}}

\newcommand{\definedas}{\equiv}
\newcommand{\defineterm}[1]{\textit{#1}}

\newcommand{\diffd}{\mathrm{d}}          
\newcommand{\diff}[1]{\diffd#1\,}        

\newcommand{\days}{\mathrm{d}}
\renewcommand{\sec}{\mathrm{s}}
\newcommand{\hours}{\mathrm{h}}
\newcommand{\Hz}{\mathrm{Hz}}
\newcommand{\mHz}{\mathrm{mHz}}

\newcommand{\pInfo}{\mathcal{I}}
\newcommand{\prob}[2]{P\left(#1\middle|#2,\pInfo\right)}
\newcommand{\probI}[1]{P\left(#1\middle|\pInfo\right)}
\newcommand{\scalar}[2]{\left(#1\middle|#2\right)}

\newcommand{\detVec}[1]{\mbf{#1}}

\newcommand{\prior}[1]{\MakeLowercase{#1}}

\newcommand{\segidx}{{\ell}} 
\newcommand{\maxseg}{m} 
\newcommand{\maxsegp}{\mathbbm{m}} 
\newcommand{\maxsegpX}{\mathbbm{m}(X)} 

\newcommand{\sumX}{\sum\limits_{X=1}^{\Ndet}}  
\newcommand{\sumXc}{\sum\limits_X}  
\newcommand{\sumSeg}{\sum\limits_{\segidx=1}^{\Nseg}}  
\newcommand{\sumSegc}{\sum\limits_{\segidx}}  
\newcommand{\sumSegci}{\sum_{\segidx}}  
\newcommand{\sumXSeg}{\sum\limits_{X\segidx}}  

\newcommand{\prodSegci}{\prod_{\segidx}}  

\newcommand{\avgSeg}[1]{\overline{#1}}                  

\newcommand{\Dop}{\lambda}

\newcommand{\Freq}{f}
\newcommand{\fdot}{{\dot{\Freq}}}

\newcommand{\Amp}{\mathcal{A}}
\newcommand{\cosi}{\cos\iota}
\newcommand{\phio}{\phi_0}

\newcommand{\Sn}{S}     
\newcommand{\SnX}{\Sn^X}                
\newcommand{\SnXal}{\Sn^X_\alpha}       

\newcommand{\sqrtSnH}{\sqrt{\Sn^{\LHO}}}             
\newcommand{\sqrtSnL}{\sqrt{\Sn^{\LLO}}}             

\newcommand{\dVx}{\detVec{x}}
\newcommand{\dVy}{\detVec{y}}
\newcommand{\dVn}{\detVec{n}}
\newcommand{\dVh}{\detVec{h}}

\newcommand{\Hyp}{\mathcal{H}}
\newcommand{\Gauss}{\mathrm{\MakeUppercase{G}}}
\newcommand{\Signal}{{\mathrm{\MakeUppercase{S}}}}
\newcommand{\Line}{{\mathrm{\MakeUppercase{L}}}}
\newcommand{\HypS}{\Hyp_\Signal}
\newcommand{\HypL}{\Hyp_\Line}
\newcommand{\HypG}{\Hyp_\Gauss}

\providecommand{\sc}[1]{\widehat{#1}}
\renewcommand{\sc}[1]{\widehat{#1}}
\newcommand{\coh}[1]{\widetilde{#1}}
\newcommand{\HypSsc}{\sc{\Hyp}_{\Signal}}
\newcommand{\HypLsc}{\sc{\Hyp}_{\Line}}
\newcommand{\HypGsc}{\sc{\Hyp}_{\Gauss}}
\newcommand{\HypGLsc}{\sc{\Hyp}_{\Gauss\Line}}

\newcommand{\Bayes}{B}

\newcommand{\OSGLsc}{\sc{O}_{{\Signal/\Gauss\Line}}} 
\newcommand{\OSGsc}{\sc{O}_{{\Signal/\Gauss}}}       

\newcommand{\OLGsc}{\sc{O}_{{\Line/\Gauss}}}
\newcommand{\BSGsc}{\sc{\Bayes}_{\Signal/\Gauss}}
\newcommand{\BSLsc}{\sc{\Bayes}_{\Signal/\Line}}
\newcommand{\BSGLsc}{\sc{\Bayes}_{\Signal/\Gauss\Line}}

\newcommand{\oLGsc}{\prior{\OLGsc}}
\newcommand{\oLGscX}{\oLGsc^X}
\newcommand{\oSGsc}{\prior{\OSGsc}}

\newcommand{\lineprobsc}{\sc{p}_{{\Line}}}
\newcommand{\lineprobscX}{\sc{p}_{{\Line}}^X}

\newcommand{\F}{\mathcal{F}}            

\newcommand{\Ftho}{\cohF_*^{(0)}}
\newcommand{\Fthop}{\cohF_*^{(0)\prime}}

\newcommand{\cohF}{\coh{\F}}

\newcommand{\scF}{\sc{\F}}
\newcommand{\scFX}{\scF^X}

\newcommand{\avF}{\avgSeg{\F}}

\newcommand{\avFpv}{\avgSeg{\F_{\mathrm{pv}}}}

\newcommand{\pDet}{p_{\mathrm{det}}}    

\newcommand{\AND}{\;\mathrm{and}\;}
\newcommand{\OR}{\;\mathrm{or}\;}

\newcommand{\Ndet}{{N_{\mathrm{det}}}}
\newcommand{\Nseg}{{N_{\mathrm{seg}}}}
\newcommand{\NsegX}[1]{{N_{\mathrm{seg}}^{#1}}}
\newcommand{\Tseg}{{T_{\mathrm{seg}}}}

\newcommand{\eto}{\ensuremath{\mathrm{e}^}}     

\newcommand{\Psft}{\mathcal{P}_{\sft}} 
\newcommand{\PsftXk}{\mathcal{P}_{\sft}^{X\segidx}}
\newcommand{\PsftXkargs}[2]{\mathcal{P}_{\sft}^{#1#2}}

\newcommand{\sft}{{\mathrm{SFT}}}
\newcommand{\Tsft}{T_\sft}

\newcommand{\Nsft}{N_\sft}

\newcommand{\LHO}{\mathrm{H1}}

\newcommand{\LLO}{\mathrm{L1}}

\newcommand{\tstart}{t_{\mathrm{start}}}
\newcommand{\tend}{t_{\mathrm{end}}}

\newcommand{\Tobs}{T_{\mathrm{obs}}}

\newcommand{\Transient}{{\mathrm{t}}}
\newcommand{\Transline}{{\mathrm{\Transient\Line}}}
\newcommand{\Transsig}{{\mathrm{\Transient\Signal}}}
\newcommand{\HypTL}{\sc{\Hyp}_{\Transline}}
\newcommand{\HypTS}{\sc{\Hyp}_{\Transsig}}
\newcommand{\HypTSk}{\coh{\Hyp}_{\Transsig}^{\segidx}}
\newcommand{\HypSTS}{\sc{\Hyp}_{\Signal\Transsig}}
\newcommand{\HypTLXk}{\coh{\Hyp}_{\Transline}^{X\segidx}}
\newcommand{\HypTLXkargs}[2]{\coh{\Hyp}_{\Transline}^{#1#2}}
\newcommand{\HypGk}{\coh{\Hyp}_{\Gauss}^{\segidx}}
\newcommand{\HypGXk}{\coh{\Hyp}_{\Gauss}^{X\segidx}}
\newcommand{\HypGkarg}[1]{\coh{\Hyp}_{\Gauss}^{#1}}
\newcommand{\HypGXkargs}[2]{\coh{\Hyp}_{\Gauss}^{#1#2}}

\newcommand{\OSGLTL}{\sc{O}_{\Signal/\Gauss\Line\Transline}}
\newcommand{\BSGLTL}{\sc{\Bayes}_{\Signal/\Gauss\Line\Transline}}
\newcommand{\BSGLTLlo}{\sc{\Bayes}_{\Signal/\Gauss\Line\Transline,\mathrm{lo}}}
\newcommand{\BTSGLTL}{\sc{\Bayes}_{\Transsig/\Gauss\Line\Transline}}
\newcommand{\OTSGLTL}{\sc{O}_{\Transsig/\Gauss\Line\Transline}}
\newcommand{\OSTSGLTL}{\sc{O}_{\Signal\Transsig/\Gauss\Line\Transline}}
\newcommand{\BSTSGLTL}{\sc{\Bayes}_{\Signal\Transsig/\Gauss\Line\Transline}}
\newcommand{\BSTSGLTLlo}{\sc{\Bayes}_{\Signal\Transsig/\Gauss\Line\Transline,\mathrm{lo}}}
\newcommand{\oTLGsc}{\sc{o}_{\Transline/\Gauss}}
\newcommand{\oTLGXk}{\coh{o}_{\Transline/\Gauss}^{X\segidx}}
\newcommand{\oTLGXkargs}[2]{\coh{o}_{\Transline/\Gauss}^{#1#2}}
\newcommand{\HypLTL}{\sc{\Hyp}_{\Line\Transline}}
\newcommand{\HypGLTL}{\sc{\Hyp}_{\Gauss\Line\Transline}}
\newcommand{\oTSGk}{\coh{o}_{\Transsig/\Gauss}^{\segidx}}
\newcommand{\oTSGsc}{\sc{o}_{\Transsig/\Gauss}}

\newcommand{\oLTLGsc}{\sc{o}_{\Line\Transline/\Gauss}}

\newcommand{\condpriorprob}{p} 
\newcommand{\cppS}{\sc{\condpriorprob}_{\Signal}}
\newcommand{\cppTS}{\coh{\condpriorprob}_{\Transsig}}
\newcommand{\cppTSk}{\coh{\condpriorprob}_{\Transsig}^{\segidx}}
\newcommand{\cppTSkarg}[1]{\coh{\condpriorprob}_{\Transsig}^{#1}}
\newcommand{\cppLX}{\sc{\condpriorprob}_{\Line}^X}
\newcommand{\cppTLXk}{\coh{\condpriorprob}_{\Transline}^{X\segidx}}
\newcommand{\cppTLXkargs}[2]{\coh{\condpriorprob}_{\Transline}^{#1#2}}

\newcommand{\cppLTLsc}{\sc{p}_{\Line\Transline}}

\newcommand{\Tinj}{{T_{\mathrm{inj}}}}

\newcommand{\hoscaled}{h_0/\sqrt{\Sn}}

\newcommand{\hoscaledthr}{{h_0^{90\%}/\sqrt{\Sn}}}

\newcommand{\denomterm}{\sc{D}}
\newcommand{\denomtermset}{\mathcal{\denomterm}}
\newcommand{\denommax}{\denomterm_{\mathrm{max}}}
\newcommand{\numerterm}{\sc{E}}
\newcommand{\numertermset}{\mathcal{\numerterm}}
\newcommand{\numermax}{\numerterm_{\mathrm{max}}}

\newcommand{\Nsky}{N_{\mathrm{sky}}}
\newcommand{\deltafdot}{\delta\fdot}
\newcommand{\Deltafdot}{\Delta\fdot}
\newcommand{\deltafreq}{\delta\Freq}
\newcommand{\Deltafreq}{\Delta\Freq}
\newcommand{\Ntemplates}{N_{\mathrm{templ}}}
\newcommand{\refinementFactor}{\gamma_{\mathrm{r}}}

\newcommand{\probBigl}[2]{P\left(\left.#1\,\right|\,#2\right)}

\newcommand{\swsq}{\mathsmaller{\square}}
\newcommand{\sbsq}{\mathsmaller{\blacksquare}}

\newcommand{\dcc}{LIGO-P1500159}

\newcommand{\eref}[1]{\eqref{#1}}
\newcommand{\fref}[1]{Fig.~\ref{#1}}

\defcitealias{keitel2014:_linerobust}{Paper~I}
\newcommand{\PaperI}{\citetalias{keitel2014:_linerobust}}

\makeatletter
\long\def\@caption#1[#2]#3{\par\begingroup
    \@parboxrestore
    \normalsize
    \@makecaption{\csname fnum@#1\endcsname}{\ignorespaces #3}\par
  \endgroup}
\long\def\@makecaption#1#2{\vskip \abovecaptionskip 
 \begin{lessindented}
 \item[]{\bf #1.} #2
 \end{lessindented}\vskip\belowcaptionskip}
\newenvironment{lessindented}{\begin{lessindented}}{\end{lessindented}}
\def\lessindented{\list{}{\itemsep=0\p@\labelsep=0\p@\itemindent=0\p@
   \labelwidth=0\p@\leftmargin=20\p@\topsep=0\p@\partopsep=0\p@
   \parsep=0\p@\listparindent=15\p@}\footnotesize\rm}

\makeatother

\begin{document}

\title[Transients in semicoherent CW searches]
{Robust semicoherent searches for continuous gravitational waves\\
with noise and signal models including hours to days long transients}

\author{David~Keitel}
\email{david.keitel@ligo.org}
\affiliation{Albert-Einstein-Institut, Callinstra{\ss}e 38, 30167 Hannover, Germany}
\affiliation{Universitat de les Illes Balears, IAC3---IEEC, 07122 Palma de Mallorca, Spain}

\vspace{10pt}

\date{24 April 2016 \\
..LIGO document number: \dcc{}}


\begin{abstract}
 The vulnerability to single-detector instrumental artifacts in
 standard detection methods for long-duration quasimonochromatic gravitational waves
 from nonaxisymmetric rotating neutron stars (`continuous waves', \defineterm{CWs})
 was addressed in
 past work~[Keitel, Prix, Papa, Leaci and Siddiqi, Phys.~Rev.~D \textbf{89}, 064023 (2014)]
 by a Bayesian approach.
 An explicit model of persistent single-detector disturbances
 led to a generalized detection statistic
 with improved robustness against such artifacts.
 Since many strong outliers in semicoherent searches of LIGO data
 are caused by transient disturbances that
 last only a few hours,
 we extend the noise model
 to cover such limited-duration disturbances,
 and demonstrate increased robustness in realistic simulated data.
 Besides long-duration CWs,
 neutron stars could also emit transient signals which,
 for a limited time,
 also follow the CW signal model (\defineterm{tCWs}).
 As a pragmatic alternative to specialized transient searches,
 we demonstrate how to make standard semicoherent CW searches
 more sensitive to transient signals.
 Considering tCWs
 in a single segment of a semicoherent search,
 Bayesian model selection yields a new detection statistic
 that does not add significant computational cost.
 On simulated data,
 we find that it increases sensitivity towards tCWs,
 even of varying durations,
 while not sacrificing sensitivity to classical CW signals,
 and still being robust to transient
 or persistent single-detector instrumental artifacts.
\footnote{\textit{
This is the author's version of an article
[Phys. Rev. D 93, 084024, doi:10.1103/PhysRevD.93.084024]
published by the American Physical Society
under the terms of the Creative Commons Attribution 3.0 License.
Further distribution of this work must maintain
attributionto the author(s) and
the published article’s title, journal citation, and DOI.
}
}
\vspace*{3\baselineskip}
\end{abstract}

\maketitle

\section{Introduction}
\label{sec:introduction}
Among the main search targets of terrestrial interferometric detectors~\cite{LIGORef:2009,GEORef:2010,
VirgoRef:2011,aasi2014:_aligo,acernese2014:_adVIRGO} are continuous gravitational waves (CWs):
periodic, narrow-band signals with a slow frequency evolution, emitted by rotating neutron stars with
nonaxisymmetric deformations.~\cite{prix06:_cw_review,JMcD2013:_maxelastic}
Searches for CWs from unknown sources over wide parameter spaces are usually performed with
\defineterm{semicoherent} methods.~\cite{brady2000:_hierarchical,cgk2005:_stackslide,krishnan04:_hough,
pletsch2009:_gct}
For these, the data are split into several \defineterm{segments},
each spanning part of the observation time.
Each segment is analyzed coherently,
and the resulting per-segment detection statistics are combined incoherently, e.g., by a sum.
At fixed computational cost, semicoherent methods are generally more sensitive than fully coherent
searches.~\cite{brady2000:_hierarchical,cgk2005:_stackslide,prix12:_optimal}

Even though gravitational-wave (GW) detectors are highly precise instruments,
they still produce complicated data sets with many noise components.
These are not all fully modeled by existing data analysis procedures,
and thus result in \defineterm{outliers} of the detection statistics.
To distinguish noise outliers from real signals,
a lot of work is routinely invested in detailed investigation of search results and auxiliary data.
Some of it can be saved by modifying detection methods
to produce less outliers in the first place.~\cite{keitel2014:_linerobust}

Many outliers in CW searches are caused by so-called \defineterm{lines},
narrow-band disturbances that are typically present for a sizable fraction of the observation time.
Such \defineterm{persistent} lines can have diverse instrumental or environmental origins,
such as harmonics of the electrical power grid frequency,
of the detector's suspension system, or from digital components.~\cite{christensen2010:_S6detchar,
coughlin2010:_lineidentification, Aasi2012:_virgochar,Accadia2012:_noemi,aasi13:_eathS5,aasi2014:_S6detchar}

A separate class of noise artifacts are \defineterm{transient `glitches'}~\cite{Blackburn2008:_glitch,
Aasi2012:_virgochar, Slutsky2010:_falsealarms,prestegard2012:_transartifact}
lasting only for a few milliseconds or seconds.
These are mostly relevant in searches for
transient GWs from compact-binary coalescences and `burst'-type events.
However, there is a third class of intermediate `transient' disturbances:
they are much longer than glitches,
so that they are highly relevant for CW searches;
but still much shorter than the full observation time,
so that they are not easily addressed by methods for mitigating persistent lines.
Typical time scales range from less than an hour to at most a few days.
\footnote{These time scales are called 'very long' in Ref.~\cite{thrane2015:_longstoch},
compared to the classical ms--s 'bursts';
but are merely 'medium' compared to the months or years spanned by CW searches.}

Such medium-duration transients,
of a linelike quasimonochromatic type,
were noticed in a semicoherent search
for CWs from the Galactic center with two years of LIGO S5 data
\cite{aasi2013:_gc-search,behnke2014:_gcmethods,behnke2013:_phdthesis},
based on the matched-filter \mbox{\defineterm{$\F$-statistic}}~\cite{jks98:_data,cutler05:_gen_fstat}
and the global-correlations (GCT) semicoherent search method~\cite{pletsch08:_global_corr,pletsch2009:_gct}.
In that search, many strong outliers could be traced back to narrow-band disturbances
in the data happening only within a single segment (each 11.5 hours long) of data
from a single detector.
Similar transient single-segment disturbances have also been found in LIGO S6 data,
using 60-hour segments for one year of data.~\cite{piccinni2014:_thesis}

In the Galactic-center search a \defineterm{permanence veto} was introduced~\cite{aasi2013:_gc-search,
behnke2014:_gcmethods,behnke2013:_phdthesis}
to remove any candidates for which a single segment contributed excessively to the semicoherent
\emph{multi-detector} detection statistic.
It was proven to be very effective, and also safe regarding classical CW signals,
which are persistent over time scales comparable to
the full length of the data set.~\cite{aasi2013:_gc-search,behnke2014:_gcmethods,
behnke2013:_phdthesis}
However, in a semicoherent CW search over several months of data,
such a veto also suppresses nonpersistent signals with durations similar to a segment length,
i.e., only a few hours or days:
these would produce just the same data signature as a disturbance in terms of single-segment,
\emph{multi-detector} statistics.
Such `\defineterm{transient-CW}' signals (\defineterm{tCWs}),
following the standard CW signal model but for a limited duration,
are also considered viable candidates for detection~\cite{prix11:_transient,nacho2014:_thesis},
with several possible emission mechanisms from perturbed neutron stars~\cite{lyne2000:_glitches,
eysden2008:_gravradglitches,bennett2010:_cwglitchrecovery,levin2011:_fmodes,kashiyama2011:_magnetars,
singh2016:_ekman}.

Therefore, this paper investigates an alternative approach to the permanence veto,
constructing a detection statistic that is robust against \emph{single-detector} transient artifacts,
while at the same time being more sensitive than standard statistics
to transient signals that are coherent across multiple detectors.

Two methods for detecting medium-duration transient signals
have been previously proposed.
One approach is to extend a coherent $\F$-statistic-based CW search
to the case of tCWs by including their duration, start-time and shape
as free parameters in the search grid.~\cite{prix11:_transient,nacho2014:_thesis}
This is nearly optimal in the Neyman-Pearson sense~\cite{neyman33:_efficient},
but computationally expensive due to the increased dimensionality of the search space.
\footnote{See Appendix A.3 of Ref.~\cite{prix11:_transient} for computing cost estimates.}
Alternatively, an unmodeled `excess power' detection method originally used to search for GW bursts of at
most a few seconds has recently been extended to cover longer durations.~\cite{thrane2015:_longstoch}
While not specifically aimed at or optimized for tCWs,
it could also be sensitive to this type of signal.
Due to the different signal models, data processing and search methods
of the CW-based and burst-based approaches,
direct comparison of their tCW sensitivity is a difficult open question;
and neither of these approaches has yet been used for an analysis of actual interferometer data.

In contrast, the approach in this paper
is a pragmatic extension of the \defineterm{line-robust statistics}
of Ref.~\cite{keitel2014:_linerobust} (hereafter~\PaperI),
which in turn are based on the standard matched-filter $\F$-statistic~\cite{jks98:_data,cutler05:_gen_fstat}.
The $\F$-statistic is close to optimal as a detection statistic for persistent CWs in Gaussian
noise~\cite{prix09:_bstat}, which in current detectors is a good model for the noise distribution over most of
the observation time and frequency range.
(See, e.g., Refs.~\cite{abbott2004:_geoligo,aasi13:_eathS5,behnke2013:_phdthesis}.)
In fact, the $\F$-statistic corresponds to a binary hypothesis test between a CW signal hypothesis and a
Gaussian-noise hypothesis.~\cite{prix09:_bstat}

In the line-robust statistics from \PaperI,
the noise model is extended to include persistent single-detector lines,
without any detailed physical modeling of the lines' origin:
the idea is to simply model a line as identical to a CW signal, but confined to a single detector.
We summarize these developments in Sec.~\ref{sec:recap}.

In Sec.~\ref{sec:LTRS} of this paper,
the new material begins with a further extension of the noise model
that also includes transient disturbances
-- or, more specifically, single-segment, single-detector disturbances.
With this approach, CW searches now become more \emph{robust} towards both persistent and transient
single-detector disturbances.
It can reproduce the robustness of the permanence veto when considering persistent CW signals only, while not
being as strict in suppressing transient tCW signals.

A second step, described in Sec.~\ref{sec:transsig}, aims to
\emph{improve} the sensitivity of semicoherent $\F$-statistic-based searches towards transient signals,
hence reducing the need for more specialized tCW searches.
We achieve this by also including an explicit signal model for transient CW-like signals on a time-scale
corresponding to a single segment in a semicoherent search.
We then test these extended detection statistics in Sec.~\ref{sec:tests}, using simulated data with a
realistic distribution of gaps in observation times, and conclude in Sec.~\ref{sec:conclusions}.
A comparison to other search methods for medium-duration transient GW
signals~\cite{prix11:_transient,nacho2014:_thesis,thrane2015:_longstoch}
remains a topic for further work.

\vspace{2.5\baselineskip}

\section{Summary of existing semicoherent detection statistics}
\label{sec:recap}
This section briefly summarizes previous work
on how the matched-filter $\F$-statistic~\cite{jks98:_data,cutler05:_gen_fstat}
follows from Bayesian hypothesis testing~\cite{prix09:_bstat},
on the permanence veto~\cite{aasi2013:_gc-search,behnke2014:_gcmethods,behnke2013:_phdthesis},
and on \PaperI's extension of the Bayesian approach to produce line-robust detection statistics.

This section also serves as an introduction to the notation used in this paper.
$x^X(t)$ denotes a time series of GW strain measured in a detector $X$.
Following the multi-detector notation of~\cite{cutler05:_gen_fstat,prix06:_searc},
boldface indicates a multi-detector vector,
i.e., $\dVx(t)$ is the multi-detector data vector with components $x^X(t)$.

For Bayesian hypothesis testing~\cite{jaynes:_logic_of_science},
$\prob{\Hyp}{x}$ is the probability of a hypothesis $\Hyp$ given data $x$ and prior information $\pInfo$.
Posterior odds ratios between two hypotheses $\Hyp_A$, $\Hyp_B$ are written with an uppercase symbol
$O_{A/B}(x,\pInfo)$; if $\Hyp_B$ is the logical sum of two hypotheses,
\mbox{$\Hyp_B=\left(\Hyp_C \OR \Hyp_D\right)$},
we write \mbox{$O_{A/B}(x,\pInfo)=O_{A/CD}(x,\pInfo)$}.
The corresponding prior odds take a lowercase symbol, $o_{A/B}(\pInfo)$, and the likelihood ratio or
\defineterm{Bayes factor} is $\Bayes_{A/B}(x,\pInfo)$, so that
\mbox{$O_{A/B}(x,\pInfo) = o_{A/B}(\pInfo) \, \Bayes_{A/B}(x,\pInfo)$}.

Also, in this paper semicoherent quantities from a search with a number $\Nseg$ of segments
carry a hat, such as $\scF$.
Coherent single-segment quantities have a tilde above the symbol and an upper index
\mbox{$\segidx=1 \dots \Nseg$} enumerating the segments, such as $\cohF^\segidx$.


\subsection{The \texorpdfstring{$\scF$}{F}-statistic: signals in Gaussian noise}
\label{sec:recap-Fstat}
We start with a \defineterm{Gaussian-noise hypothesis},
\mbox{$\HypGsc: \dVx(t) = \dVn(t)$},
with the samples of $\dVn(t)$ drawn from a Gaussian distribution.
Its posterior probability, given priors $\probI{\HypGsc}$ and $\probI{\dVx}$, is
\begin{equation}
  \label{eq:pHG}
  \prob{\HypGsc}{\dVx} = \frac{\probI{\HypGsc}}{\probI{\dVx}}\,\kappa\,\eto{-\frac{1}{2}\scalar{\dVx}{\dVx}} \,,
\end{equation}
with a normalization constant $\kappa$ and a scalar product between time series defined as
\begin{equation}
  \label{eq:scalarproduct}
  \scalar{\dVx}{\dVy} \equiv \sumX \frac{1}{\SnX}\int_0^T x^X(t)\,y^X(t)\,\diff{t}\,.
\end{equation}
Here, $\SnX$ are the single-sided power-spectral densities (PSDs),
assumed as uncorrelated between different detectors
and constant over the (narrow) frequency band of interest.

The \defineterm{CW signal hypothesis}
\begin{equation}
 \label{eq:CWmodel}
 \HypSsc: \dVx(t) = \dVn(t) + \dVh(t;\Amp,\Dop)
\end{equation}
contains a waveform with a set $\Amp$ of four \defineterm{amplitude parameters} and a set $\Dop$ of
\defineterm{phase-evolution parameters}
(including frequency, spin-down
and sky position).
In a semicoherent search, different $\Amp^\segidx$ are allowed in each segment; but we simplify our notation 
by redefining \mbox{$\Amp=\left\{ \Amp^\segidx \right\}$}.

After marginalizing over $\Amp$ and the associated prior
(cf.~\cite{prix09:_bstat,prix11:_transient,keitel2014:_linerobust,keitelprix2015:_LRSsafety}),
the posterior probability is
\begin{equation}
 \label{eq:pHS}
 \hspace{-0.75em}
  \prob{\HypSsc}{\dVx} = \oSGsc(\pInfo)\,\prob{\HypGsc}{\dVx}\,\eto{\scF(\dVx) - \Nseg\Ftho}\,.
\end{equation}
Here,
\mbox{$\oSGsc(\pInfo) \equiv \probI{\HypS} \left/ \probI{\HypG} \right.$}
are the prior odds between a signal and Gaussian noise;
$\Ftho \in (-\infty,\infty)$ is a free parameter (the result of an arbitrary $\Amp$-prior cutoff);
and the semicoherent multi-detector $\scF$-statistic is,
for a single parameter-space point $\Dop$,
given by the sum over single-segment coherent $\cohF^\segidx$-statistics:
\begin{equation}
  \label{eq:scFstat}
  \scF(\dVx;\Dop) \equiv \sum\limits_{\segidx=1}^{\Nseg} \cohF^\segidx(\dVx^\segidx;\Dop)\,.
\end{equation}
In practice, often an \defineterm{interpolating StackSlide} algorithm is used, where $\scF(\dVx;\Dop)$ for each
$\Dop$ is computed from a set of $\cohF^\segidx(\dVx^\segidx;\Dop^\segidx)$ with the $\Dop^\segidx$
picked from a coarser grid in parameter space than the
$\Dop$.~\cite{brady2000:_hierarchical,cgk2005:_stackslide,prix12:_optimal,pletsch2009:_gct}.
In Eq.~\eref{eq:pHS}, as well as for the rest of the paper, we do not explicitly show the $\Dop$-dependence of our
detection statistics.

These posterior probabilities can be used to compute odds ratios between the different hypotheses.
First, we see from Eqs.~\eref{eq:pHG} and \eref{eq:pHS} that
\begin{equation}
 \label{eq:OSG}
 \OSGsc(\dVx,\pInfo) \equiv \frac{\prob{\HypSsc}{\dVx}}{\prob{\HypGsc}{\dVx}} \propto \BSGsc(\dVx,\pInfo) \propto \eto{\scF(\dVx)} \,,
\end{equation}
i.e., this Bayesian approach reproduces the $\scF$-statistic as
the Neyman-Pearson-optimal detection statistic
for CW signals in pure Gaussian noise and under the assumed priors.
The free parameter $\Ftho$ is irrelevant in this case.

\subsection{Permanence veto}
\label{sec:recap-pveto}
The permanence veto, as introduced in
Refs.~\cite{aasi2013:_gc-search,behnke2014:_gcmethods,behnke2013:_phdthesis},
works by the following algorithm:
From a fixed Gaussian false-alarm level or some real-data noise studies, a threshold $\avF_{\mathrm{thr}}$ is
set on the average semicoherent statistic $\avF\definedas\scF/\Nseg$.
Then, for each candidate the highest single-segment contribution is removed, defining
\begin{equation}
  \label{eq:avFstat-pveto}
  \avFpv(\dVx;\Dop) \equiv
  \frac{1}{\Nseg-1} \sum\limits_{\segidx \neq \maxseg} \cohF^\segidx(\dVx^\segidx;\Dop^\segidx)\,.
\end{equation}
where $\maxseg$ is the segment with the highest multi-detector
\mbox{$\cohF^{\maxseg}\definedas\max_{\segidx}\cohF^{\segidx}$}.
\vspace{0.5\baselineskip}

In the original implementation of Refs.~\cite{aasi2013:_gc-search,behnke2014:_gcmethods,behnke2013:_phdthesis},
the $\avFpv$ value of each candidate is compared with the threshold $\avF_{\mathrm{thr}}$ to determine
whether to veto the candidate.
In our tests in Sec.~\ref{sec:tests}, we slightly modify this algorithm to treat the permanence veto on a more
equal footing with the other detection statistics:
We define $\avFpv$ exactly as in Eq.~\eref{eq:avFstat-pveto}, but we set a detection threshold by computing
the maximum of $\avFpv$ over a pure-noise data set.

\subsection{Line-robust statistics}
\label{sec:recap-LRS}
\PaperI{} introduced a more general noise model including
a simple noncoincident \defineterm{`line' hypothesis}
\mbox{$\HypLsc^{X} :  x^{X}(t) = n^{X}(t) + h^{X}(t;\Amp^{X})$},
which just assumes a CW-like disturbance in an arbitrary single detector $X$.
It leads to a line-robust detection statistic which is reproduced here with slightly updated notation.

Marginalisation as for Eq.~\eref{eq:pHS} yields
\begin{equation}
  \label{eq:pHL}
  \prob{\HypLsc}{\dVx} = \prob{\HypGsc}{\dVx} \, \sumX \oLGscX \, \eto{\scFX(x^X) - \Nseg\Ftho}\,,
\end{equation}
with the per-detector line-prior odds and their sum,
\begin{subequations}
 \begin{align}
  \oLGscX(\pInfo) &\equiv {\probI{\HypLsc^X}}/{\probI{\HypGsc^X}} \,,\\
  \oLGsc(\pInfo)  &\equiv \sumXc \oLGscX(\pInfo) \,.
 \end{align}
\end{subequations}
We suppress the $\pInfo$-dependence of any odds ratios or Bayes factors in Eq.~\eref{eq:pHL} and from now on.

Furthermore, we can combine the (mutually exclusive) hypotheses $\HypGsc$ and $\HypLsc$ into an extended noise
hypothesis \mbox{$\HypGLsc \equiv ( \HypGsc \OR \HypLsc )$}, with posterior probability

\begin{align}
 \label{eq:pHGL}
  &\prob{\HypGLsc}{\dVx} = \prob{\HypGsc}{\dVx} + \prob{\HypLsc}{\dVx} \\
  &= \prob{\HypGsc}{\dVx} \left( 1 + \sumX \oLGscX \eto{\scFX(x^X) - \Nseg\Ftho} \nonumber
\right) \,.
\end{align}

Finally, using Eqs.~\eref{eq:pHS} and \eref{eq:pHGL},
we obtain generalized signal-versus-noise odds
\begin{equation}
  \label{eq:OSGL}
  \OSGLsc(\dVx)
  = \frac{\oSGsc \, \eto{\scF(\dVx)}}
         {\eto{\Nseg\Ftho} + \sumXc \, \oLGscX \eto{\scFX(x^X)}}
\end{equation}
and, with the conditional probabilities for lines in the absence of a signal,
\begin{subequations}
 \label{eq:lineprobs}
\begin{align}
 \label{eq:lineprobX}
 \lineprobsc  &\equiv \prob{\HypLsc}{\HypGLsc} = \frac{\oLGsc}{1 + \oLGsc} \,, \\
 \label{eq:lineprobsum}
 \lineprobscX &\equiv \prob{\HypLsc^X}{\HypGLsc} = \frac{\oLGscX}{1 + \oLGsc} \,,
\end{align}
\end{subequations}
the corresponding Bayes factor, or \defineterm{line-robust statistic}, is

\begin{equation}
 \label{eq:BSGL}
  \BSGLsc(\dVx)
  = \frac{\eto{\scF(\dVx)}}
         {(1-\lineprobsc) \, \eto{\Nseg\Ftho} + \sumXc \, \lineprobscX \eto{\scFX(x^X) }} \,.
\end{equation}
\vspace{0.5\baselineskip}

In this statistic, the parameter $\Ftho$
determines a transition scale between increased strictness to either Gaussian noise or lines.
It can therefore be considered as a tuning parameter for the line-robust statistic.
In section~VI.B of \PaperI{} it was suggested to choose the lowest $\Ftho$
that makes $\BSGLsc$ as efficient as $\scF$ for simulated CW signals in quiet (almost-Gaussian) data,
and demonstrated that this tuning choice at the same time offers improved robustness against lines.

The limit of \mbox{$\Ftho \rightarrow -\infty$} corresponds to a binary test of $\HypSsc$ against $\HypLsc$,
excluding Gaussian noise.
We refer to this Bayes factor $\BSLsc$ as the \defineterm{pure line-veto statistic}.

\section{Deriving a CW detection statistic that is robust to single-segment disturbances}
\label{sec:LTRS}
Going beyond the noise model of \PaperI,
we now turn to the issue of noncoincident transient linelike disturbances.
To address it in the same Bayesian framework as above, consider a new \defineterm{`transient-line' hypothesis}
$\HypTLXk$ for a quasiharmonic disturbance in a single segment $\segidx$ and single detector $X$:

\begin{equation}
 \label{eq:HypTLXk}
 \HypTLXk :  x^{X\segidx}(t) = n^{X\segidx}(t) + h^{X\segidx}(t;\Amp^{X\segidx}) \,.
\end{equation}
\vspace{0.25\baselineskip}

This is just the full CW hypothesis from Eq.~\eref{eq:CWmodel} restricted to a subset $x^{X\segidx}(t)$ of
the data.
Thus, in analogy with Eqs.~\eref{eq:pHS} and \eref{eq:pHL} and dropping the time-series arguments again, the
posterior probability for $\HypTLXk$ is

\begin{equation}
 \label{eq:pHTLXk}
 \prob{\HypTLXk}{x^{X\segidx}}
 = \prob{\HypGXk}{x^{X\segidx}} \, \oTLGXk \, \eto{\cohF^{X\segidx}(x^{X\segidx}) - \Ftho}\,.
\end{equation}
\vspace{0.25\baselineskip}

In principle, we could now build up a wide range of composite hypotheses about the whole data-set $\dVx$,
spanning $\Nseg \times \Ndet$ subsets $x^{X\segidx}(t)$,
by combining instances of $\HypTLXk$ and of the single-segment Gaussian-noise hypothesis $\HypGXk$,
and by setting appropriate constraints on the amplitude parameters $\{\Amp^{X\segidx}\}$.
\vspace{0.25\baselineskip}

For example,
the hypothesis $\HypLsc$ for persistent single-detector lines corresponds to
$\prodSegci\HypTLXk$ with the same $\Amp^\mathrm{\mathrm{Y}\segidx}$ for all $\segidx$,
but only for a specific detector $X=\mathrm{Y}$;
combined with $\prodSegci\HypGXk$ for all other detectors $X\neq\mathrm{Y}$.
\vspace{0.25\baselineskip}

However, we concentrate on one specific \emph{new} full-data-set hypothesis $\HypTL$:
for the case of a transient, single-detector disturbance in only one $\segidx$ and one $X$,
with no prior constraint on the values of these indices.
For example, if we have data in two segments for two detectors, the full hypothesis is

\begin{align}
 \HypTL : \quad &\left( \HypTLXkargs{1}{1} \AND \HypGXkargs{1}{2} \AND \HypGXkargs{2}{1} \AND
 \HypGXkargs{2}{2} \right) \label{eq:HypTL}\\
 \OR &\left( \HypGXkargs{1}{1} \AND \HypTLXkargs{1}{2} \AND \HypGXkargs{2}{1} \AND
 \HypGXkargs{2}{2} \right) \nonumber\\
 \OR &\left( \HypGXkargs{1}{1} \AND \HypGXkargs{1}{2} \AND \HypTLXkargs{2}{1} \AND
 \HypGXkargs{2}{2} \right) \nonumber\\
 \OR &\left( \HypGXkargs{1}{1} \AND \HypGXkargs{1}{2} \AND \HypGXkargs{2}{1} \AND
 \HypTLXkargs{2}{2} \right) \,. \nonumber
\end{align}
\vspace{0.25\baselineskip}

The full semicoherent posterior probability for this hypothesis is then

\vspace*{-0.5\baselineskip}
\begin{align}
 \prob{\HypTL}{\dVx} &= \sumXSeg \prob{\HypTLXk}{x^{X\segidx}}
        \hspace{-0.5em} \underset{\OR \segidx' \neq \segidx}{\underset{Y \neq X}{\prod }}
        \hspace{-0.5em} \prob{\HypGXkargs{Y}{\segidx'}}{x^{Y\segidx'}} \nonumber \\
                     &= \prob{\HypGsc}{\dVx} \, \sumXSeg \oTLGXk \, \eto{\cohF^{X\segidx} - \Ftho} \,, \hspace{-2em}
                        \label{eq:pTL}
\end{align}
introducing the shorthand notation \mbox{$\sumXSeg\definedas\sumSeg\sumX$}.

We can then produce a combined noise hypothesis $\HypGLTL$ that allows for either pure Gaussian noise, a
persistent line or a single-segment transient line:
\begin{equation}
 \label{eq:HypGLTL}
 \HypGLTL : \left( \HypGsc \OR \HypLsc \OR \HypTL \right) \,.
\end{equation}
As seen before in \PaperI, $\HypL^X(\Amp^X)$ has the same likelihood as
$\HypG^X$ in the special case of vanishing amplitude parameters, \mbox{$\Amp^X=0$}.
But when we obtain the full line hypothesis $\HypL^X$ by marginalizing over $\Amp^X$, this is only a null-set
contribution; furthermore, the two hypotheses are still, by construction, \emph{logically} exclusive.
The same reasoning applies to $\HypTLXk$.
Hence, the probabilities of these three hypotheses must simply add up:
\begin{multline}
 \label{eq:pGLTL}
 \prob{\HypGLTL}{\dVx} \\
                      \,\,\, = \prob{\HypGsc}{\dVx} + \prob{\HypLsc}{\dVx} + \prob{\HypTL}{\dVx} \\
                       = \prob{\HypGsc}{\dVx} \,
                          \left( 1 + \sumX \oLGscX \, \eto{\scFX - \Nseg\Ftho} \right. \\
                           + \left. \sumXSeg \oTLGXk \, \eto{\cohF^{X\segidx} - \Ftho} \right) \,.
\end{multline}

Then, the odds ratio between the classical persistent-CW signal hypothesis $\HypSsc$ and the
combined triple-noise hypothesis $\HypGLTL$
yields a new detection statistic
\begin{multline}
 \label{eq:OSGLTL}
 \OSGLTL = \oSGsc \, \eto{\scF} \left/ \left( \eto{\Nseg\Ftho}
                                              + \sumX \oLGscX \eto{\scFX} \right. \right. \\
 \left. + \sumXSeg \oTLGXk \eto{\cohF^{X\segidx}+(\Nseg-1)\Ftho} \right) \,,
\end{multline}
where, just as a reminder, the semicoherent $\scF$-statistics are \mbox{$\scF = \sumSegci\cohF^{\segidx}$}
and \mbox{$\scFX=\sumSegci\cohF^{X\segidx}$}.

With the total prior disturbance odds \mbox{$\oLGsc \equiv \sumXc \oLGscX$} and
\mbox{$\oTLGsc \equiv \sumXSeg \oTLGXk$}, we introduce the following shorthands for prior probabilities
conditional on the composite noise hypothesis $\HypGLTL$,
generalizing the $\lineprobsc$ and $\lineprobscX$ from Eq.~\eref{eq:lineprobs}:
\vspace{-0.5\baselineskip}
\begin{subequations}
\label{eq:priorweights}
\begin{align}
 \label{eq:lineweights}
 \cppLX &\definedas \prob{\HypL^X}{\HypGLTL}
        =          \frac{\oLGscX}{1+\oLGsc+\oTLGsc} \,, \\
 \label{eq:translineweights}
 \cppTLXk &\definedas \prob{\HypTL^{X\segidx}}{\HypGLTL}
          =          \frac{\oTLGXk}{1+\oLGsc+\oTLGsc} \,, \\
 \label{eq:LTLprob}
 \cppLTLsc &\definedas \prob{\HypLTL}{\HypGLTL}
          = \frac{\oLTLGsc}{1 + \oLTLGsc} \,.
\end{align}
\end{subequations}
This allows us to write the corresponding Bayes factor as
\begin{widetext}
\begin{equation}
 \label{eq:BSGLTL}
 \BSGLTL = \frac{\eto{\scF}}
           {(1-\cppLTLsc) \, \eto{\Nseg\Ftho} + \sumXc \cppLX \, \eto{\scFX}
           + \sumXSeg \cppTLXk \, \eto{\cohF^{X\segidx} + (\Nseg-1)\Ftho}}  \,.
\end{equation}
\end{widetext}

We see that the difference between
(i) the persistent-line term already present in the $\BSGLsc$ of Ref.~\eref{eq:BSGL} and
(ii) the newly introduced transient-line term
is that we have either
(i) a sum over $X$ of the exponentials of a sum over $\segidx$ of $\cohF^{X\segidx}$, or
(ii) a double sum over $X$ and $\segidx$ of the exponentials of each individual $\cohF^{X\segidx}$ plus a
large constant term $(\Nseg-1)\Ftho$.

Hence, if there is a strong disturbance in a single $(X, \segidx)$ combination and if the transition-scale
parameter $\Ftho$ has been chosen as higher than the typical $\cohF^{X\segidx}$ in pure Gaussian noise (in
accordance with the tuning procedure described in Sec.~VIB of \PaperI), then the
latter term can dominate in the denominator.
This will make $\BSGLTL$ stricter in suppressing these transient disturbances than $\BSGLsc$.

We could have introduced an additional free tuning parameter into $\BSGLTL$ by using a different cutoff on the
$\Amp^{X\segidx}$ prior in $\HypTLXk$ than for the $\Amp^X$ in $\HypLsc^X$, resulting in a different $\Fthop$
appearing. 
However, we already have freedom in the relative weights of persistent and transient-line contributions
through the $\cppLX$ and $\cppTLXk$, and there is no clear physical motivation in such a complication of the
amplitude priors (which were chosen \textit{ad hoc}, to reproduce the $\F$-statistic, in the first place, cf.
Refs.~\cite{prix09:_bstat,prix11:_transient}).
Hence, we refrain from this possibility, and use the tests in Sec.~\ref{sec:tests} to demonstrate sufficient
flexibility of $\BSGLTL$ without it.

As the denominator of $\BSGLTL$ is a sum of exponentials (or weighted exponentials, but of course the
log of the weights can be absorbed into the exponents), it is often dominated by a single term.
The same is true for $\BSGLsc$, and its limiting behavior in various cases was discussed in Sec.~IVB1 of
\PaperI.
Here, we just give an expression for $\ln\BSGLTL$ written as a sum of the dominant term and a logarithmic
correction,

\begin{equation}
  \label{eq:lnBSGLTL}
  \ln\BSGLTL = \scF - \denommax
               - \ln \left( \sum\limits_{\denomterm \in \denomtermset} \eto{\denomterm-\denommax} \right)\,,
\end{equation}
\vspace{0.5\baselineskip}

\noindent where $\denommax\definedas\max\denomtermset$ is the maximum of the set of exponents,
with \mbox{$1+\Ndet(1+\Nseg)$} elements:
\begin{equation}
\begin{split}
 \label{eq:denomtermset}
 \denomtermset \definedas &\left\{
                                   \Nseg\Ftho + \ln\left( 1-\cppLTLsc \right) ,\,
                                   \scFX + \ln\cppLX , \right. \\
                          &\left.\hphantom{\}\Nseg} \cohF^{X\segidx} + (\Nseg-1)\Ftho + \ln\cppTLXk
                           \right\} \,.
\end{split}
\end{equation}

In computer implementations, this form is useful both for numerical stability (avoiding underflows)
and to speed up computation when the correction term can be neglected,
\mbox{$\ln\BSGLTL \approx \scF - \denommax$}.

We also consider an intermediate step where we reduce the $\sumXSeg$-sum in the denominator to the highest
per-segment contributions from each detector, but keep the remaining $1+2\Ndet$ terms.
This will reduce computational cost while also corresponding to the initial assumption of a single-segment
disturbance:
again, because of the exponentials, a single significantly increased $\cohF^{X\segidx}$ will easily dominate
over all others.
Hence, in many cases a good approximation to the Bayes factor is given by
\begin{align}
 \label{eq:BSGLTL_loudestonly}
 &\BSGLTL \approx \eto{\scF} \left/ \left(
                   \left( 1-\cppLTLsc \right) \, \eto{\Nseg\Ftho} \right. \right. \\
 &\left. + \sumXc \cppLX \, \eto{\scFX}
         + \sumXc \cppTLXkargs{X}{\maxsegpX} \, \eto{\cohF^{X\maxsegpX}
         + (\Nseg-1)\Ftho} \right) \,, \nonumber
\end{align}
\vspace{0.5\baselineskip}

\noindent where $\maxsegpX$ is the segment number for which
\mbox{$\cppTLXkargs{X}{\maxsegpX}\,\eto{\cohF^{X\maxsegpX}}
       \definedas \max_{\segidx}\left(\cppTLXk\,\eto{\cohF^{X\segidx}}\right)$}.

\vspace{0.5\baselineskip}

In some applications,
purely for reasons of search code simplification and reduction of data volume,
only reduced single-segment information may be available:
the set of values \mbox{$\{\cohF^{\maxseg},\{\cohF^{X\maxseg}\}\}$}
only for the segment $\maxseg$ with the highest \emph{multi-detector}
\mbox{$\cohF^{\maxseg}\definedas\max_{\segidx}\cohF^{\segidx}$}.
To still obtain an approximate version of $\BSGLTL$ in such cases,
we define a modified \defineterm{`loudest-only' detection statistic}
\begin{align}
 \label{eq:BSGLTL_loudestonly2}
 &\BSGLTLlo \definedas \eto{\scF} \left/ \left(
                           \left( 1-\cppLTLsc \right) \, \eto{\Nseg\Ftho} \right. \right. \\
 & \left. + \sumX \cppLX \, \eto{\scFX}
          + \sumX \cppTLXkargs{X}{\maxseg} \, \eto{\cohF^{X\maxseg}
          + (\Nseg-1)\Ftho} \right)  \,. \nonumber
\end{align}
This quantity could, in principle,
differ quite significantly from the actual Bayes factor $\BSGLTL$.
There is also no guarantee that it is as efficient a detection statistic
under our initial hypotheses,
so we will test its efficiency with simulated data in Sec.~\ref{sec:tests}.

\vspace{2\baselineskip}

\section{Deriving a detection statistic for persistent or transient signals, robust to persistent or transient
single-detector lines}
\label{sec:transsig}
CW-like transient signals might be interesting search
targets.~\cite{prix11:_transient,nacho2014:_thesis,singh2016:_ekman}
One might now anticipate that the transient-line-robust Bayes factor
$\BSGLTL$ from Eq.~\eqref{eq:BSGLTL}
is \emph{too restrictive} towards these,
as a multi-detector-coherent signal in a single segment
can increase the denominator of Eq.~\eref{eq:BSGLTL} more than the numerator.

However, the approach of considering more general hypotheses built up from the set
$\{\cohF^{\segidx},\{\cohF^{X\segidx}\}\}$ should actually allow for \emph{more sensitivity} towards
transient signals than any detection statistic based only on the total semicoherent results, like $\scF$
and $\OSGLsc$.

So we try to improve over $\BSGLTL$ by deriving another generalized detection statistic,
answering the following question:
how likely is \emph{any type of CW-like signal} (persistent or transient), in comparison with
Gaussian noise, a persistent line, or a transient line?

Starting from the full set of single-segment $\{\cohF^{\segidx},\{\cohF^{X\segidx}\}\}$, the most
general answer would involve a large set of hypotheses for signals in different numbers of segments.
But here we keep to the simplifying assumption of single-segment transients, introducing a
transient-signal hypothesis as the multi-detector version of Eq.~\eref{eq:HypTLXk}:
\begin{equation}
 \label{eq:HypTSk}
 \HypTSk :  \dVx^{\segidx}(t) = \dVn^{\segidx}(t) + \dVh^{\segidx}(t;\Amp^{\segidx}) \,.
\end{equation}
Note that this is different from the single-segment, single-detector transient-line hypothesis $\HypTLXk$
from Eq.~\eref{eq:HypTLXk} only if the data set for segment $\segidx$ contains data for at least two detectors $X$.
In this section, we assume this to be the case for the whole data set.
However, in the real world the components of a multi-detector network often have different duty factors and
standard data selection methods~\cite{shaltev2013:_thesis} can result in segments with data
from one detector only, or with negligible amounts of data from the other detectors.

We test the robustness of this detection statistic,
derived with the assumption of full segment coverage by all detectors,
by considering a data set with realistic duty factors in Sec.~\ref{sec:tests},
and discuss ways to deal with the slight issues it can cause
in Sec.~\ref{sec:tests-TS}.

Let us continue from the posterior distribution for $\HypTSk$, which is analogous to Eq.~\eref{eq:pHTLXk}:

\begin{equation}
 \label{eq:pHTSk}
 \prob{\HypTSk}{\dVx^{\segidx}}
 = \prob{\HypGk}{\dVx^{\segidx}} \, \oTSGk \, \eto{\cohF^{\segidx}(\dVx^{\segidx}) - \Ftho}\,.
\end{equation}
\vspace{0.25\baselineskip}

\noindent The hypothesis $\HypTS$ for a transient signal in an arbitrary segment is
the logical OR combination of $\HypTSk$ analogous to Eq.~\eref{eq:HypTL},
so that the posterior $\prob{\HypTS}{\dVx}$ is obtained in analogy with
Eq.~\eref{eq:pTL}:
\begin{align}
 \prob{\HypTS}{\dVx} &= \sumSeg \prob{\HypTSk}{\dVx^{\segidx}}
                        \prod_{\segidx' \neq \segidx} \prob{\HypGkarg{\segidx'}}{\dVx^{\segidx'}} \nonumber\\
                     &= \prob{\HypGsc}{\dVx} \, \sumSeg \oTSGk \eto{\cohF^{\segidx} - \Ftho} \,. \label{eq:pTS}
\end{align}
\vspace{0.5\baselineskip}

Testing for tCW signals only, this yields an odds ratio

\begin{align}
 \label{eq:OTSGLTL}
 &\OTSGLTL \hspace{-0.1em}=\hspace{-0.1em} \sumSegc \oTSGk \, \eto{\cohF^{\segidx}+(\Nseg-1)\Ftho} \hspace{-0.4em} \\
 &\left/ \left( \eto{\Nseg\Ftho} \hspace{-0.2em}+ \sumXc \oLGscX \, \eto{\scFX}
               \hspace{-0.2em}+ \sumXSeg \oTLGXk \, \eto{\cohF^{X\segidx}(\Nseg-1)\Ftho}\hspace{-0.1em}\right)
   \right.\hspace{-0.2em}. \nonumber
\end{align}

Just as for the various noise hypotheses,
we can also add up the probabilities for the signal hypotheses $\HypSsc$ and $\HypTS$
to evaluate a more general persistent-or-transient `CW-like' hypothesis:
\begin{align}
 \label{eq:pSTS}
 &\prob{\HypSTS}{\dVx} = \prob{\HypSsc}{\dVx} + \prob{\HypTS}{\dVx} \\
 &= \prob{\HypGsc}{\dVx} \left( \oSGsc \, \eto{\scF - \Nseg\Ftho}
    + \sumSeg \oTSGk \, \eto{\cohF^{\segidx} - \Ftho} \right)\hspace{-0.2em}. \nonumber
\end{align}

The odds ratio between generalized signal hypothesis and generalized noise hypothesis is then
\begin{align}
 \label{eq:OSTSGLTL}
 &\OSTSGLTL \hspace{-0.1em}=\hspace{-0.1em} \left(\hspace{-0.1em}\oSGsc \, \eto{\scF}
            \hspace{-0.2em}+ \sumSegc \oTSGk \, \eto{\cohF^{\segidx}+(\Nseg-1)\Ftho}
            \hspace{-0.1em}\right) \hspace{-0.4em} \\
 &\left/ \left( \eto{\Nseg\Ftho} \hspace{-0.2em}+ \sumXc \oLGscX \, \eto{\scFX}
               \hspace{-0.2em}+ \sumXSeg \oTLGXk \, \eto{\cohF^{X\segidx}(\Nseg-1)\Ftho}\hspace{-0.1em}\right)
   \right.\hspace{-0.2em}. \nonumber
\end{align}

The corresponding generalized Bayes factor follows by
introducing additional prior-weight variables
in analogy to $\cppLX$, $\cppTLXk$
from Eq.~\eref{eq:priorweights}:
\begin{align}
 \label{eq:signalweights}
 \hspace{-1cm}
 \cppS &\definedas \prob{\HypS}{\HypSTS}
         =          \frac{\oSGsc}{\oSGsc+\oTSGsc} \nonumber \\
        &=  \left( 1 - \cppTS \right)
         = \left( 1 - \sumSeg \cppTSk \right)
\end{align}

for persistent signals and
\begin{equation}
 \label{eq:transsigweights}
 \hspace{-1.2cm}
 \cppTSk \definedas \prob{\HypTSk}{\HypSTS}
          =          \frac{\oTSGk}{\oSGsc+\oTSGsc}
\end{equation}
for transient signals.
\vfill

This persistent-or-transient `CW-like' robust detection statistic is then given by
\begin{widetext}
\begin{equation}
 \label{eq:BSTSGLTL}
 \BSTSGLTL = \frac{ (1-\cppTS) \, \eto{\scF}
             + \sumSegc \cppTSk \, \eto{\cohF^{\segidx}+(\Nseg-1)\Ftho} }
             { (1-\cppLTLsc) \, \eto{\Nseg\Ftho} + \sumXc \cppLX \, \eto{\scFX}
                   + \sumXSeg \cppTLXk \, \eto{\cohF^{X\segidx} + (\Nseg-1)\Ftho}}  \,.
\end{equation}
\end{widetext}

As was the case for $\BSGLTL$ from Eq.~\eref{eq:BSGLTL}, additional freedom in tuning this statistic could be
obtained from different amplitude-prior cutoffs in $\HypLsc$, $\HypTL$ and now also $\HypSsc$ and $\HypTS$.
But again we restrict ourselves to using the same cutoff, resulting in a single tuning parameter $\Ftho$, and
use only the set of prior variables $\{\cppS,\cppTSk,\cppLTLsc,\cppTLXk\}$ as weights for the various
contributions.

Next, we consider the logarithm of this Bayes factor, splitting numerator and denominator separately into sums
of a dominant term and a logarithmic correction, which generalizes Eq.~\eref{eq:lnBSGLTL}:

\begin{align}
  \label{eq:lnBSTSGLTL}
  \ln\BSGLTL =&\;\hphantom{-} \numermax
               + \ln \left( \sum\limits_{\numerterm \in \numertermset} \eto{\numerterm-\numermax} \right) \\
              &- \denommax
               - \ln \left( \sum\limits_{\denomterm \in \denomtermset} \eto{\denomterm-\denommax} \right)\,,
               \nonumber
\end{align}
\vspace{0.5\baselineskip}

where $\denommax$ is the maximum of the same set of denominator exponents given in Eq.~\eref{eq:denomtermset}
and $\numermax=\max\numertermset$ is the maximum of the numerator exponents:

\begin{equation}
 \label{eq:numertermset}
 \numertermset = \left\{
                         \scF + \ln\cppS , \;
                         \cohF^{\segidx} + (\Nseg-1)\Ftho + \ln\cppTSk
                 \right\} \,.
\end{equation}

For transient signals and disturbances
that are indeed limited to a single segment
(or reasonably close),
it should suffice to compute an approximate Bayes factor using only the maximum single-segment
contributions:

\begin{gather}
 \label{eq:BSTSGLTL_loudestonly}
 \BSTSGLTL \approx \left( \cppS \, \eto{\scF}
                           + \cppTSkarg{\maxsegp} \, \eto{\cohF^{\maxsegp}+(\Nseg-1)\Ftho}
                   \right) \nonumber \\
 \left/ \left(
    \left( 1-\cppLTLsc \right) \, \eto{\Nseg\Ftho} + \sumXc \cppLX \, \eto{\scFX}
 \right. \right. \\
 \left.
    \hspace{3em} + \sumXc \cppTLXkargs{X}{\maxsegpX} \, \eto{\cohF^{X\maxsegpX} + (\Nseg-1)\Ftho}
 \right) \,, \nonumber
\end{gather}

\noindent where $\maxsegp$ is the segment number with the largest multi-detector contribution,
so that
\mbox{$\cppTSkarg{\maxsegp}\,\eto{\cohF^{\maxsegp}}
       \definedas \max_{\segidx}\left(\cppTSk\,\eto{\cohF^{\segidx}}\right)$},
and $\maxsegpX$ is the analogous segment number for each detector:
\mbox{$\cppTLXkargs{X}{\maxsegpX}\,\eto{\cohF^{X\maxsegpX}}
       \definedas \max_{\segidx}\left(\cppTLXk\,\eto{\cohF^{X\segidx}}\right)$}.
\vspace{0.5\baselineskip}

As in Eq.~\eref{eq:BSGLTL_loudestonly2},
we also define an \textit{ad hoc} modified `loudest-only' detection statistic
where we use only information from one segment $\maxseg$ with the highest \emph{multi-detector}
\mbox{$\cohF^{\maxseg}\definedas\max_{\segidx}\cohF^{\segidx}$}:

\begin{gather}
 \BSTSGLTLlo \definedas \left( \cppS \, \eto{\scF}
                               + \cppTSkarg{\maxseg} \, \eto{\cohF^{\maxseg}+(\Nseg-1)\Ftho}
                        \right) \nonumber \\
 \left/ \left(
    \left( 1-\cppLTLsc \right) \, \eto{\Nseg\Ftho} + \sumXc \cppLX \, \eto{\scFX}
 \right. \right. \label{eq:BSTSGLTL_loudestonly2} \\
 \left.
    \hspace{2em} + \sumXc \cppTLXkargs{X}{\maxseg} \, \eto{\cohF^{X\maxseg} + (\Nseg-1)\Ftho}
 \right) \,. \nonumber
\end{gather}

Again this requires empirical tests to verify
that it is close in efficiency to the full Bayes factor,
which will be demonstrated in Sec.~\ref{sec:tests}.

Alternatively, in a search for tCWs only,
or for CWs and tCWs with two separate toplists,
one could use the Bayes factor corresponding to Eq.~\eref{eq:OTSGLTL}:

\begin{align}
 \label{eq:BTSGLTL}
 &\BTSGLTL = \sumSegc \cppTSk \, \eto{\cohF^{\segidx}+(\Nseg-1)\Ftho}
            \left/ \left( (1-\cppLTLsc) \, \eto{\Nseg\Ftho} \right. \right.  \nonumber \\
            & \left. \hspace{0.75cm} + \sumXc \cppLX \, \eto{\scFX}
                   + \sumXSeg \cppTLXk \, \eto{\cohF^{X\segidx} + (\Nseg-1)\Ftho} \right) \,.
\end{align}

All these expressions also simplify significantly if all \mbox{$\oTSGk=\oSGsc$} and \mbox{$\oTLGXk=\oLGscX$},
which we assume for most of the test cases in the next section.

\section{Tests on simulated data}
\label{sec:tests}
In this section, we present some tests of the new Bayes factors $\BSGLTL$ and $\BSTSGLTL$ in the form of
injection studies on simulated data, where simulated CW and tCW signals (`injections') are recovered from
simulated noise.
We use the same basic injection procedure and detection criteria as described
in Sec.~VIIB of \PaperI.

\subsection{Search setup and data sets}
\label{sec:tests-setup}
For two reasons, it is important to test these detection statistics with realistic data and a search setup
that is close to what is used in practice:
First, the approach in this paper is to provide a simple extension of the established search codes
that already produce the $\scF$-statistic and line-robust statistics,
which should be directly applicable in current search efforts,
and hence tested in similar circumstances.
Second, as we are interested in transient features,
the time-domain characteristics of real data sets are important
for any performance demonstration, especially the occurrence of gaps in the data:
it is necessary to test that gaps do not lead to persistent CW signals being rejected,
or to a smaller improvement in sensitivity towards tCW signals than in perfectly continuous data.

Hence, we use fully simulated data, but with realistic duty factors taken from the real LIGO S6 data.
One data set contains pure Gaussian noise,
whereas an additional transient non-Gaussian disturbance is present in the second data set.

\vspace*{-0.5\baselineskip}
\subsubsection{Search setup}
Our search setup mimics the Einstein@Home~\cite{EatH} `S6Bucket' search~\cite{EatHS6Bucket} on LIGO S6 data:
we use data spanning about $255\,\days$, analyzed semicoherently with \mbox{$\Nseg=90$} segments of
\mbox{$\Tseg=60\,\hours$}.

The analysis is performed with the \texttt{HierarchSearchGCT} code~\cite{lalsuite},
a semicoherent StackSlide implementation based on the GCT method of Ref.~\cite{pletsch2009:_gct}.
We use the same search grids as the S6Bucket search,
covering the whole sky
and only the first-order spin-down parameter $\fdot$.
\texttt{HierarchSearchGCT} is limited to semicoherent refinement in spindown only
(by a factor $\refinementFactor$)
but not over the sky,
a limitation that has been identified
as an important point for future improvement.~\cite{manca:_gctrefinement,wette2015:_metric}

The search output is a \defineterm{toplist} of the most significant candidates
ranked by one of the semicoherent statistics $\scF$ or $\BSGLsc$.
For this study, we have modified the code to also return the single-segment
$\cohF^{\segidx}$- and $\cohF^{X\segidx}$-statistics
for each toplist candidate.

\begin{table}[b!]
 \begin{tabular}{l r}
 \hline\hline\\[-0.1cm]
  Common search parameters                & \\ \hline
  Detectors                               & LIGO $\LHO$, $\LLO$ \\
  $\tstart$ $[\sec]$                      & 949469977 \\
  $\tend$ $[\sec]$                        & 971529850 \\
  $\Nseg$, $\NsegX{\LHO}$, $\NsegX{\LLO}$ & 90, 89, 90 \\
  $\Tseg$                                 & $60\hours$ \\
  Frequency resolution $\deltafreq$       & $\approx1.6143 \times 10^{-6\hphantom{1}}\,\Hz\hphantom{^2}$ \\ 
  Spin-down resolution $\deltafdot$       & $\approx5.7890 \times 10^{-11}\,\Hz^2$ \\ 
  $\fdot$ refinement factor $\refinementFactor$ & 230 \\
  Nominal sky-grid mismatch               & 0.3 \\
  Original toplists                       & $\scF$ and $\BSLsc(\oLGscX=0.5)$ \\
  Toplist length                          & 1000 \\[0.25cm]
  Full-band search parameters             & \\ \hline
  $\min\Freq$                             & $50.0\,\Hz$ \\
  Frequency range $\Deltafreq$            & $ 0.05\,\Hz$ \\
  $\min\fdot$                             & $\approx-2.6425 \times 10^{-9}\,\Hz^2$ \\ 
  Spin-down range $\Deltafdot$            & $\approx 2.9067 \times 10^{-9}\,\Hz^2$ \\ 
  Sky points $\Nsky$                      & 707 \\
  Search jobs (sky partitions)            & 51 \\[0.25cm] 
  Purely Gaussian data $\max2\avF$        & 6.374 \\
  Transient-line data $\max2\avF$         & 11.985 \\
  Persisent-line data $\max2\avF$         & 42.246 \\[0.25cm]
  Per-injection search box                & \\ \hline
  $\Freq$ range                           & $0.001\,\Hz$ \\
  $\fdot$ range                           & $\approx2.3156 \times 10^{-10}\,\Hz^2$ \\ 
                                          & (4 coarse-grid points) \\
  Sky points                              & 10 \\[0.1cm]
 \hline\hline
 \end{tabular}
 \caption{
  \label{tbl:searchparams}
  Search parameters for pure-noise (full-band) and per-injection searches with
  \texttt{lalapps\_HierarchSearchGCT}.
 }
\end{table}

We first analyze a $50\,\mHz$ band of each simulated noise-only data set
(purely Gaussian and Gaussian + transient disturbance),
and obtain the maximum of each detection statistic over the whole sky and $\Freq$, $\fdot$ range.

Then, for a set of fixed signal strengths $h_0$,
CW or tCW signals with otherwise random parameters
are injected into the same noise realization,
and searched for again over a smaller \defineterm{search box}.
This is a subset of the original search grid containing (but usually not centered on) the injection point.
A signal is considered as detected
if the highest value from this search box exceeds
the maximum value from the pure-noise search.
\vspace{0.5\baselineskip}

The search parameters for both the full-band noise-only search and for the smaller injection search boxes
are given in Table~\ref{tbl:searchparams}.
In all test cases, 1000 signals are injected per $h_0$ value,
with a range chosen so that detection-efficiency curves are well-sampled over the whole range from 0 to 1.
The signals are drawn with random amplitude parameters $\cosi$, $\phio$, $\Psi$;
and with $\Freq$, $\fdot$ and sky position randomly distributed over the full search range
as given in Table~\ref{tbl:searchparams}.
The distribution of tCW-specific time-domain parameters is discussed
below in Secs.~\ref{sec:tests-TS}--\ref{sec:tests-TS-varduration}.
\vspace{0.5\baselineskip}

Another point where we construct our procedure in analogy with the S6Bucket search is the ranking of
candidates in the toplists kept by the \texttt{HierarchSearchGCT} code.
For each search job (51 sky partitions per noise-only search, or one search box per injection) we keep two
toplists with 1000 candidates each.
One toplist is sorted by $\scF$ and one by the pure line-veto statistic \mbox{$\BSLsc(\oLGscX=0.5)$},
which corresponds to $\BSGLsc$ in the limit of \mbox{$\Ftho \rightarrow -\infty$}.
All other detection statistics are then computed from the \emph{union} of these two toplists.
\vspace{0.5\baselineskip}

In principle, this procedure could lead to some noise outliers or some injections being missed for the
`recomputed' statistics.
However, the two toplists (classic $\scF$-statistic and pure line-veto statistic) are very `orthogonal'
in the sense that one is nearly optimal for Gaussian data and one is tuned towards strong disturbances, so
that candidates that would be significant by one of the other Bayes factors are very likely to appear in at
least one of these two toplists.
Also, tests with longer toplists have found that this approach is generally sufficient to not lose any
would-be high-significance candidates of any recomputed statistic by having them below the threshold of
both ranking statistics.

\subsubsection{Simulated data sets}
To generate our data, we used the duty factors of the H1 and L1 detectors for the data selection of the
Einstein@Home S6Bucket search on LIGO S6 data:
this gives us 6156 Short Fourier Transforms (\defineterm{SFTs}) in H1 and 5924 SFTs in L1,
each SFT 1800\,s long, with realistic gaps in between.
\vspace{0.5\baselineskip}

The data selection method~\cite{shaltev2013:_thesis} used to generate the S6Bucket segment list was
optimized for total sensitivity and did not ensure uniform duty factors over segments and detectors.
Hence, it happens to have two particularly unequal segments, where one detector contributes no or very little
data (compared to an average of 67 SFTs per segment and detectors):
segment 64 (of 90) has no data from detector H1,
and segment 76 has only four SFTs from detector L1.
\vspace{0.5\baselineskip}

The first `quiet' data set is pure simulated Gaussian noise, from the \texttt{Makefakedata\_v5}
code~\cite{lalsuite}, with the sensitivity of the two detectors being realistically slightly unequal:
the single-sided PSDs are \mbox{$\sqrtSnH=3.2591\times10^{-22}\,\Hz^{-1/2}$} and
\mbox{$\sqrtSnL=2.9182\times10^{-22}\,\Hz^{-1/2}$}.
\vspace{0.5\baselineskip}

\begin{figure}[t!bp]   
  \includegraphics[width=\columnwidth,clip]{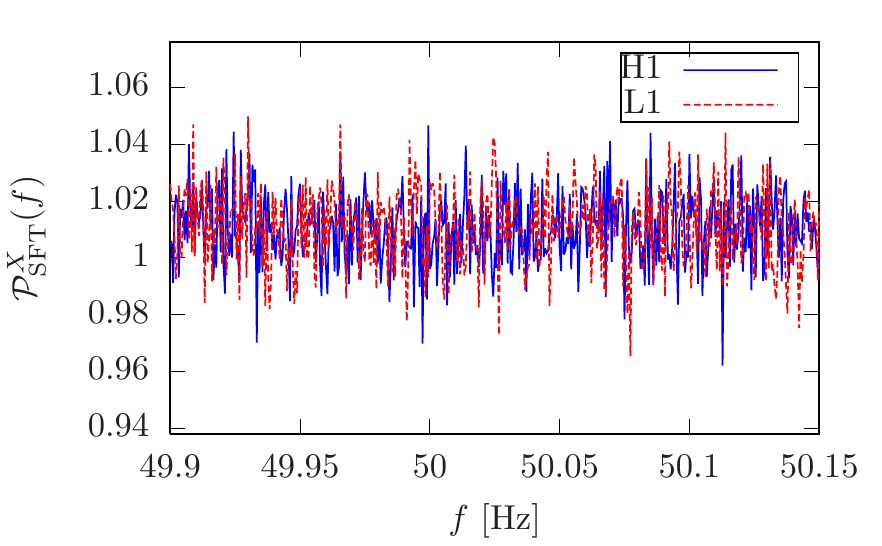} \\
  \includegraphics[width=\columnwidth,clip]{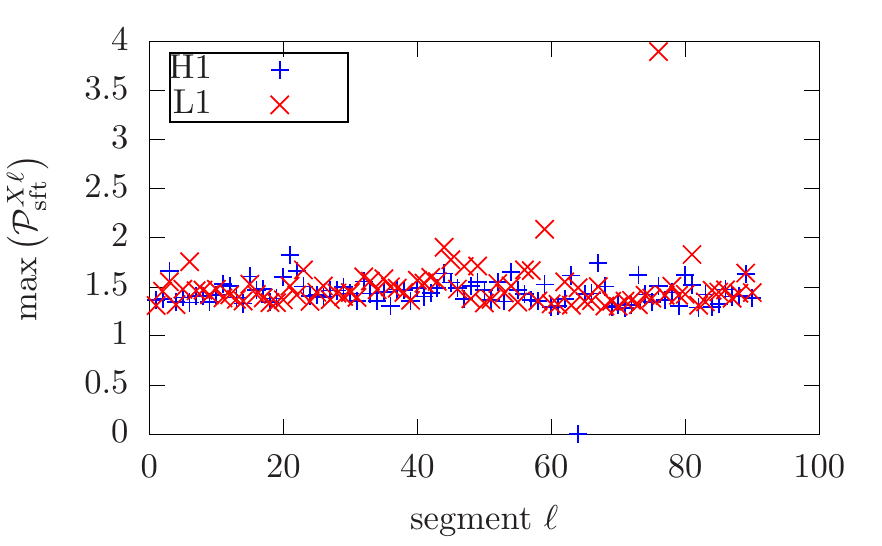}
  \caption{
   \label{fig:testbands-normSFTpower-pure-gauss}
   Pure Gaussian noise data set.
   Top panel:
   normalized SFT power $\Psft^X$ averaged over all $\Nseg=90$ segments.
   Bottom panel: 
   single-segment $\max_\Freq\PsftXk$, maximized over SFT frequency bins,
   as a function of segments $\segidx$.
  }
\end{figure}

\begin{figure}[t!bp]
  \includegraphics[width=\columnwidth,clip]{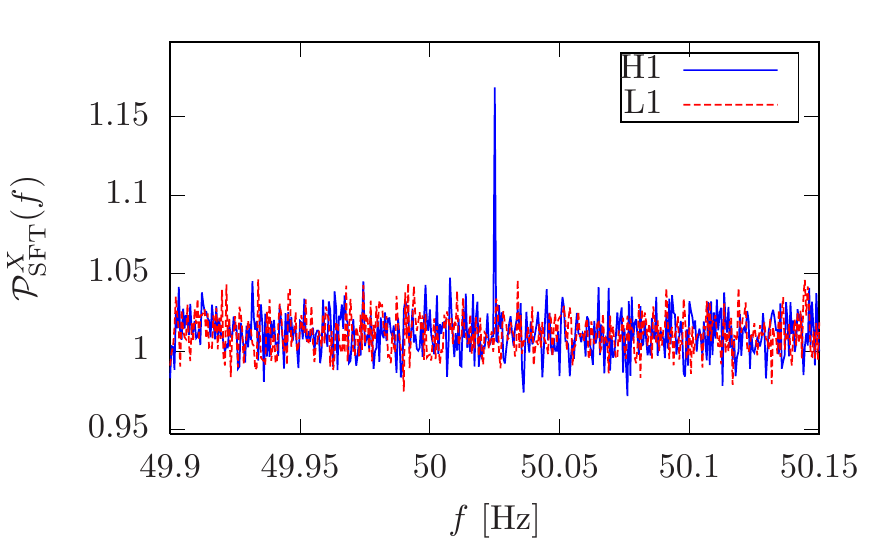} \\
  \includegraphics[width=\columnwidth,clip]{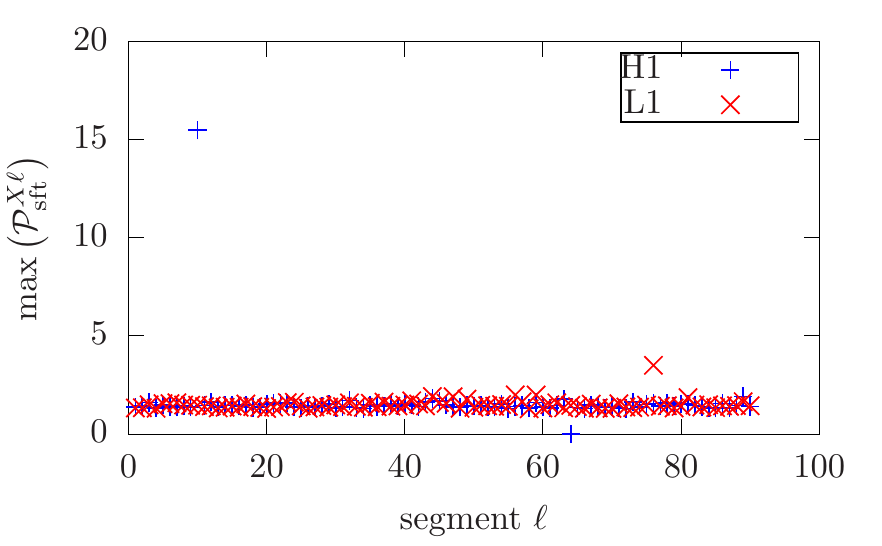}
  \caption{
    \label{fig:testbands-normSFTpower-transient-line}
    Data set with Gaussian noise and a single-detector stationary line injected for the duration of
    segment $\segidx=10$ in detector H1.
    Panels are the same as in \fref{fig:testbands-normSFTpower-pure-gauss}.
\vspace{-\baselineskip}
  }
\end{figure}

The per-detector normalized SFT power

\begin{equation}
  \label{eq:Psft}
  \Psft^X(f) \equiv \frac{2}{\Nsft\,\Tsft} \sum_{\alpha=1}^{\Nsft}
                    \frac{ \left| \widetilde{x}_\alpha^X(f)\right|^2 }{\SnXal(f)}
\end{equation}

\noindent for this data set is shown in \fref{fig:testbands-normSFTpower-pure-gauss}, both as a frequency-dependent
average $\Psft^X(f)$ over the whole data set and in the form of a single-segment maximum
$\max_\Freq\PsftXk$ over SFT frequency bins, as a function of segments $\segidx$.
The apparent outlier \mbox{$\max_\Freq\PsftXkargs{\LLO}{\segidx=76}\approx3.9$} is just an effect of
low-number statistics, as segment 76 contains only four SFTs from detector L1.

We have also generated a second data set containing a transient single-detector disturbance.
We started with an independent realization of Gaussian noise with the same time stamps and PSDs as the first
set, and then used the otherwise equivalent implementation \texttt{Makefakedata\_v4}~\cite{lalsuite}
\footnote{As of the writing of this paper, the newer MFD\_v5 code did not support stationary line injections.}
to inject a stationary line feature with
fixed amplitude \mbox{$h_{0\Line}=4\times10^{-23}$}
and frequency \mbox{$\Freq_{\Line}=50.025\,\Hz$}
in a single detector (H1) during a single segment
\mbox{$\mathrm{\segidx}_{\Line}=10$}.
A transient line in a single segment is chosen because,
as discussed in the introduction,
most strong disturbances in LIGO S5 and S6 data are indeed either persistent over the whole observation time,
or over only a single segment.\cite{behnke2014:_gcmethods,piccinni2014:_thesis}

The normalized SFT power for this data set is shown in \fref{fig:testbands-normSFTpower-transient-line}.
We see that the disturbance produces a very high single-segment
\mbox{$\max_\Freq\PsftXkargs{\LHO}{\segidx=10}\approx15$}.
It is also strong enough to show up in the average $\Psft^X(f)$ over the whole data set,
but in this average it is much weaker than the persistent lines studied before (cf. Fig.~7 of \PaperI).

\begin{figure}[t!bp]
  \includegraphics[width=\columnwidth,clip]{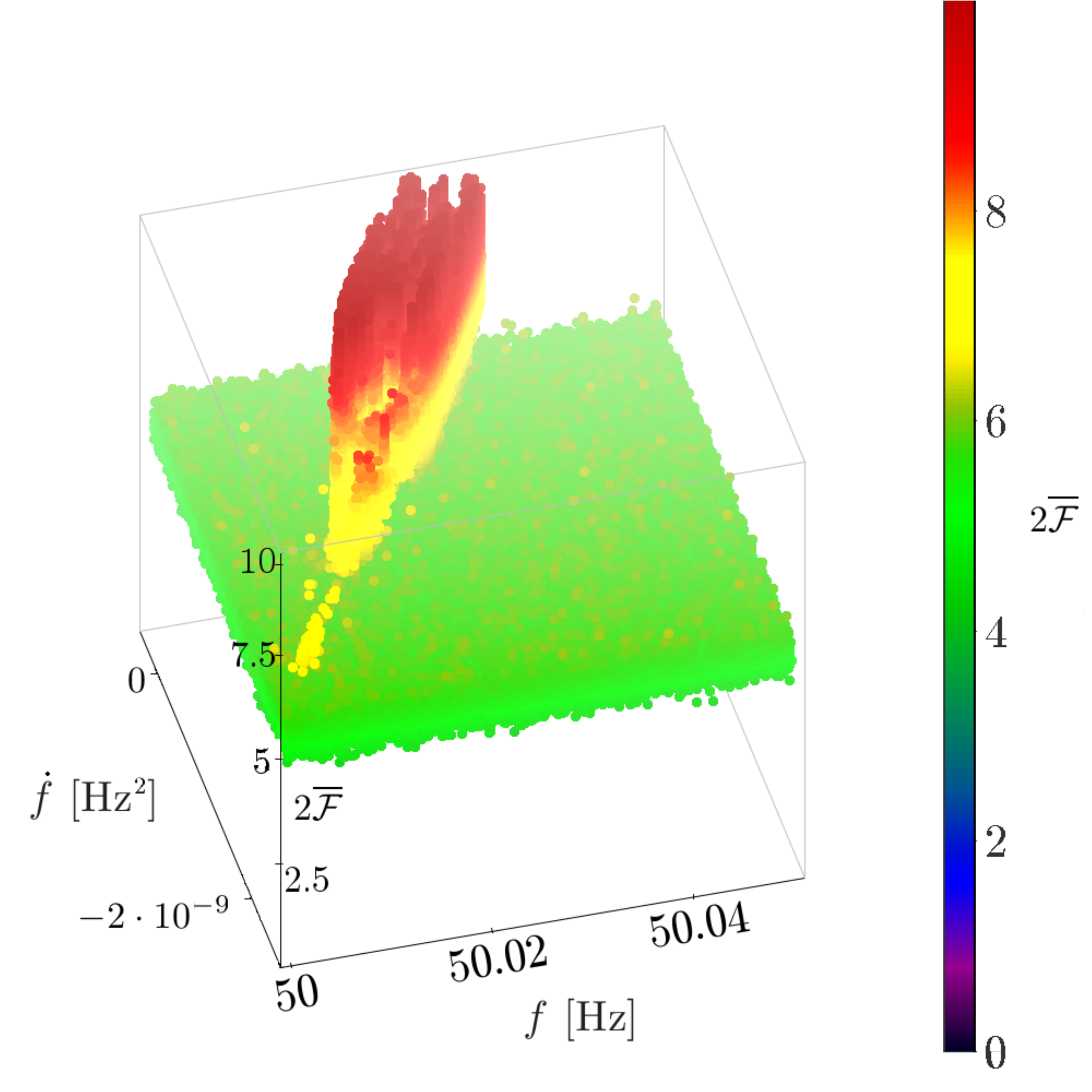}
  \caption{
    \label{fig:freq-fdot-Fstat-transient-line}
    Data set with Gaussian noise and a single-detector stationary line injected for the duration of
    segment $\segidx=10$ in detector H1.
    The figure shows the average multi-detector semicoherent $2\avF$-statistic, over 90 segments, with the
    `full-band' search parameters listed in Table~\ref{tbl:searchparams}, as a function of frequency $\Freq$
    and spin-down $\fdot$.
  }
\end{figure}

This simulated disturbance is similar to a family of transient disturbances in LIGO S6 data informally called
`pizza-slice disturbances'~\cite{piccinni2014:_thesis} due to their shape in
three-dimensional plots of $\scF$-statistics against frequency $\Freq$ and spin-down $\fdot$.
\fref{fig:freq-fdot-Fstat-transient-line} presents such a plot for our simulated data set.
Though a sharp line in $\Psft^X$, the semicoherent search sees this transient disturbance as a wide
structure in parameter space.
Different templates match the disturbance at different times, leading to the `pizza-slice' shape.
The simulation result is somewhat narrower than the typical LIGO S6 `pizza slice', since its duration is a
whole segment of \mbox{$\Tseg=60\hours$}, while the corresponding disturbances in S6 data typically last only
for a few SFTs.

We have also generated a third data set with the same procedure as the second,
but with the single-detector disturbance active over the whole observation time, i.e. as a persistent line.
In such a case, the new transient-optimized Bayes factors $\BSGLTL$ and $\BSTSGLTL$
cannot be expected to yield further improvements over
the detection efficiency of the persistent-line-robust statistic $\BSGLsc$.
Still, we have verified that in this case there are no losses either compared to $\BSGLsc$,
with both new Bayes factors reproducing the performance found for $\BSGLsc$ in \PaperI{}
and improving over the standard $\scF$-statistic.
To avoid redundancy with that paper and the purely Gaussian case (data set 1),
this set of results is not shown and discussed in detail here,
our focus being instead on the cases where improvements can be made.

\subsubsection{Tuning the free parameters of the line-robust statistics}
A global value for the transition-scale parameter $\Ftho$ is determined through requiring \defineterm{safety} in
quiet data, choosing the minimum value required to have negligible differences in detection probability to the
$\scF$-statistic.
This statement holds down to the false-alarm level probed by this study,
which is bounded by the inverse number of fine-grid templates
\mbox{$\left(\Ntemplates
 = \refinementFactor \, \Nsky \, \tfrac{\Deltafreq}{\deltafreq} \, \tfrac{\Deltafdot}{\deltafdot}
 \approx 2.3 \times 10^{11}\right)$}
in the search setup,
but is effectively somewhat higher due to template overlap.

For the original $\BSGLsc$, we re-use a value of \mbox{$\Ftho\approx3.027$}
found in more extensive studies on LIGO S6 data \cite{EatHS6Bucket}.
The present injection study on the pure Gaussian-noise data set is sampled in steps of 0.1 in $\Ftho$,
which for $\BSGLTL$ and $\BSTSGLTL$ leads to a value of \mbox{$\Ftho\approx3.0$}.
The difference is negligible with respect to the sampling accuracy of 1000 injections per $h_0$ value,
as shown by the results in the next subsection.

For the per-detector line priors,
we do not take into account our privileged knowledge from generating the data sets,
instead testing the robustness of the detection statistics by a simple choice of
\mbox{$\oLGscX=\oTLGXk=0.001$} for all $X$ and $\segidx$ in both data sets.
This corresponds to the lower truncation suggested in section~VI.A of \PaperI{}
as a conservative choice that considers lines as rare,
but still keeps the line hypothesis open in case it is strongly preferred by the data.

However, we also investigate the effect of setting \mbox{$\oTLGXkargs{X}{\segidx=64,76}=0.0$} in the two
segments with no or small contributions from one of the two detectors.
The rationale for this modification is that the single-segment signal hypothesis $\HypTSk$ of
Eq.~\eref{eq:HypTSk} becomes indistinguishable from our transient-line hypothesis $\HypTLXk$ of
Eq.~\eref{eq:HypTLXk} when that segment is completely dominated by a single detector.

For any future searches of LIGO data using these statistics,
tuning of both the transition scale $\Ftho$ and the per-detector line-priors
will be revisited using the specific search setup and data characteristics.


\subsection{Persistent CW signals}
\label{sec:tests-PS}
\begin{figure}[b!]
  \includegraphics[width=\columnwidth,clip]{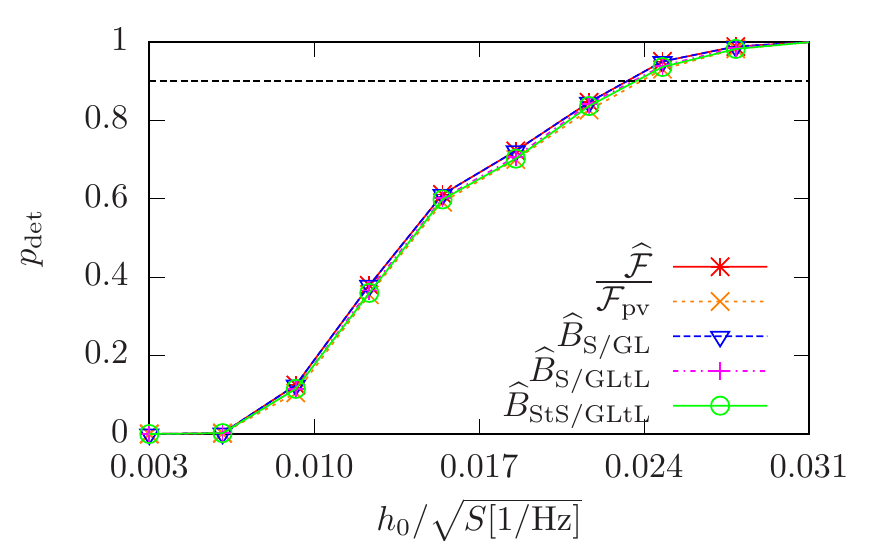} \\
  \includegraphics[width=\columnwidth,clip]{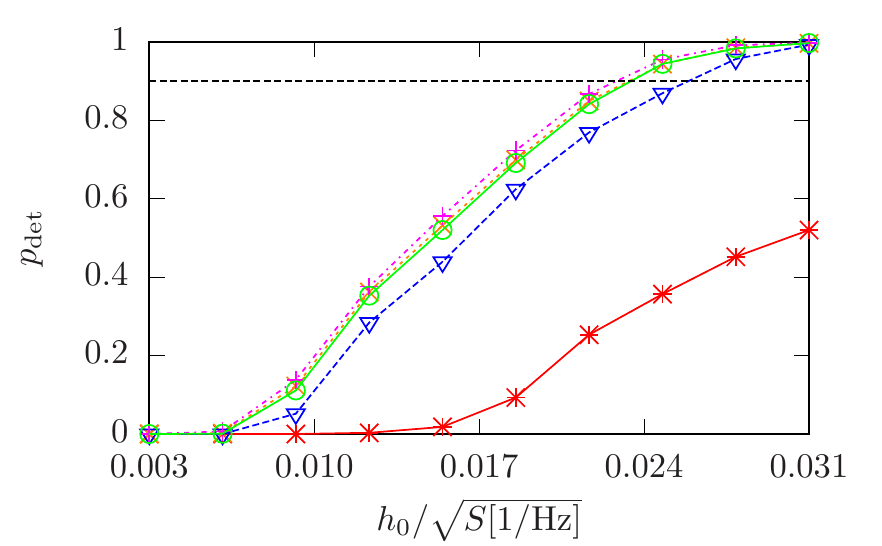}
  \caption{
    \label{fig:hsgct-simdata-detprobs-persistent}
    Detection efficiency $\pDet$ for persistent CW signals,
    as a function of scaled signal amplitude $\hoscaled$,
    for the following semicoherent statistics:
    $\scF$,
    $\avFpv$ (permanence veto),
    $\BSGLsc$ from Eq.~\eref{eq:BSGL}
    $\BSGLTL$ from Eq.~\eref{eq:BSGLTL},
    and $\BSTSGLTL$ from Eq.~\eref{eq:BSTSGLTL}.
    The dashed horizontal lines mark \mbox{$\pDet=90\%$}.
    Top    panel: injections in pure Gaussian noise.
    Bottom panel: injections in Gaussian noise with a transient disturbance.
    Statistical uncertainties are smaller than the plot markers.
  }
\end{figure}

For persistent CW signals, the injection procedure is identical to that in \PaperI.
\fref{fig:hsgct-simdata-detprobs-persistent} shows results in the form of detection probabilities $\pDet$
for the various statistics as functions of the scaled signal amplitude $\hoscaled$.

As discussed in the previous subsection,
tuning \mbox{$\Ftho=3$} allows both $\BSGLTL$ and $\BSTSGLTL$ to match
almost perfectly the detection efficiency of the $\scF$-statistic and of $\BSGLsc$ in quiet Gaussian data,
with maximum discrepancies in $\pDet$ of 1\%
(down to the false-alarm level of this search setup).
These are smaller than the statistical uncertainties from 1000 injections, and could be resolved with a more
detailed $\Ftho$ tuning.
In this case, all statistics reach 90\% detection probability at \mbox{$\hoscaledthr\approx0.023$}.

In the data set with a transient-linelike single-detector disturbance,
$\scF$ performs much worse,
while $\BSGLsc$ loses a few \% of $\pDet$ at any given $h_0$.
Here, the new $\BSGLTL$ performs best with no degradation from the quiet case, still achieving
\mbox{$\hoscaledthr\approx0.023$}.
Taking into account the possibility of tCW signals (which are not actually present in this case),
$\BSTSGLTL$ only sacrifices about 1\% in $\pDet$, and still improves significantly over $\BSGLsc$.

In both cases, using the simplified `loudest-only' detection statistics from
Eqs.~\eref{eq:BSGLTL_loudestonly2} and \eref{eq:BSTSGLTL_loudestonly2} with only one set of
single-segment $\{\cohF^{\maxseg},\{\cohF^{X\maxseg}\}\}$ values
(with $\maxseg$ chosen so that \mbox{$\cohF^{\maxseg}=\max_{\segidx}\cohF^{\segidx}$})
does in fact not decrease detection efficiency.
No extra curves are plotted for these statistics.

Also, we see that the performance of the permanence
veto~\cite{aasi2013:_gc-search,behnke2014:_gcmethods,behnke2013:_phdthesis} in the absence of tCW signals
is closely reproduced by our new Bayes factors.

\subsection{tCW signals of exactly one segment length}
\label{sec:tests-TS}
\begin{figure}[t!]
  \includegraphics[width=\columnwidth,clip]{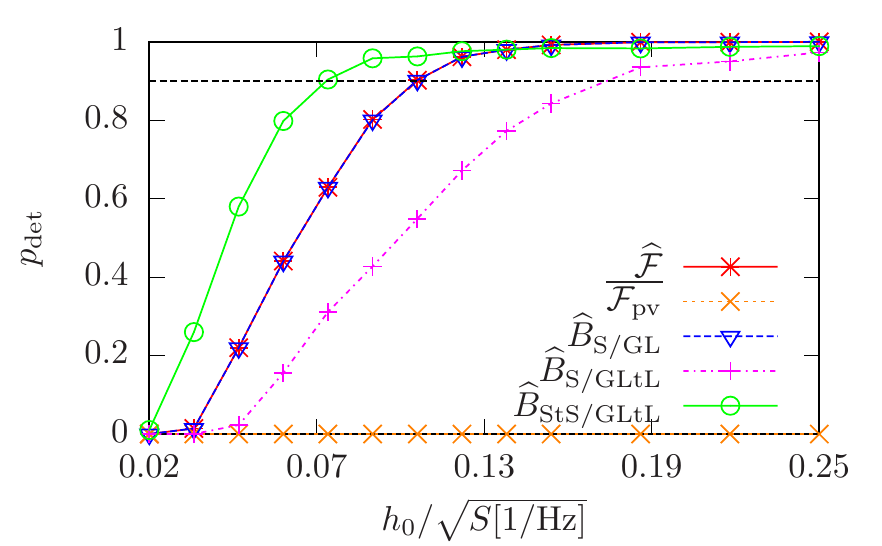} \\
  \includegraphics[width=\columnwidth,clip]{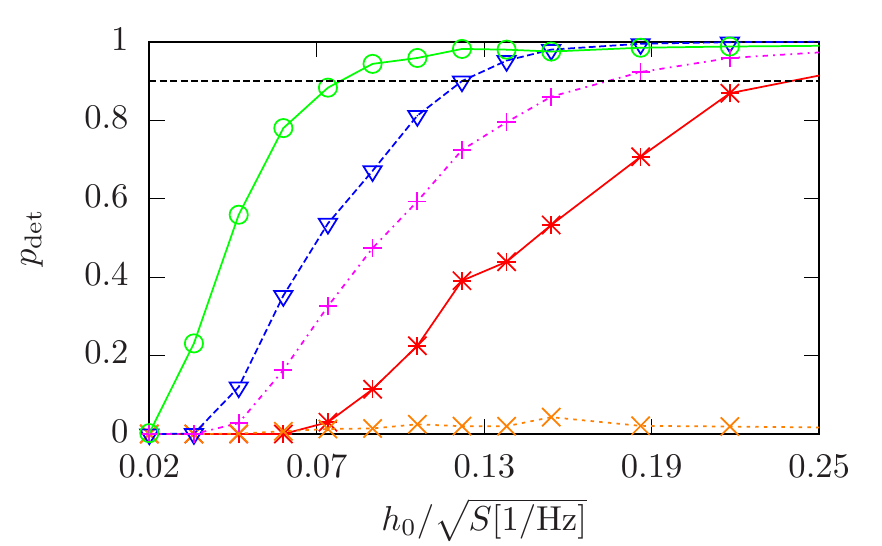}
  \caption{
    \label{fig:hsgct-simdata-detprobs-transient}
    Detection efficiency $\pDet$ for transient tCW signals
    with \mbox{$\Tinj=\Tseg=60\hours$},
    randomly distributed over segments,
    for the same statistics as in \fref{fig:hsgct-simdata-detprobs-persistent}.
    The dashed horizontal lines mark \mbox{$\pDet=90\%$}.
    Top    panel: injections in pure Gaussian noise.
    Bottom panel: injections in Gaussian noise with a transient disturbance.
  }
\end{figure}

For the first set of transient signal injections, we simulate CW-like signals that are active during exactly
one segment, i.e. with fixed \mbox{$\Tinj=\Tseg=60\hours$} and a start time corresponding to that of a
randomly picked segment for each injection.
Though not realistic,
this configuration is useful as a first test of principle,
where the assumptions made in the derivation of Sec.~\ref{sec:transsig}
correspond exactly to the data,
before generalizing the test to a more realistic signal population
with varying transient durations in the next section.
Detection probabilities for this case are shown in \fref{fig:hsgct-simdata-detprobs-transient}, over both
noise data sets (purely Gaussian and Gaussian + transient disturbance).
\vspace{0.25\baselineskip}

The established semicoherent detection statistics $\scF$ and $\BSGLsc$ achieve
\mbox{$\hoscaledthr\approx0.1$} in the first, quiet data set.
This is about a factor 4--5 worse than for persistent signals,
which is actually already a smaller ratio than expected from the naive
\mbox{$\sqrt{\tfrac{\Tobs}{\Tinj}}=\sqrt{\Nseg}$}
scaling for a fully coherent search,
but consistent with a more detailed StackSlide sensitivity estimation~\cite{wette2012:_sensitivity}.
\vspace{0.25\baselineskip}

When going from the purely Gaussian to the transient-line data set, the performance of $\scF$ and $\BSGLsc$
decreases somewhat more strongly for these tCW signals than it did for persistent signals, with $\BSGLsc$
losing up to 10\% in $\pDet$ at some $h_0$ values and increasing to \mbox{$\hoscaledthr\approx0.12$}.
\vspace{0.25\baselineskip}

Considering the permanence veto, we confirm that it would effectively remove almost all of these tCW signals,
and hence we indeed need an alternative detection statistic for this case.
\vspace{0.25\baselineskip}

The Bayes factor $\BSGLTL$, which adds to $\BSGLsc$ only the possibility of transient single-detector
disturbances (such as that in the second data set),
but not of the multi-detector-coherent transient signals we are now injecting, 
was found before to be safe for persistent CW signals.
Now it turns out to be much safer for tCWs than the permanence veto, but still performs
worse than $\BSGLsc$ in both noise data sets, with \mbox{$\hoscaledthr\approx0.17$}.
Hence, this is not a particularly safe detection statistic for tCWs.
\vspace{0.25\baselineskip}

On the other hand, the full transient-signal-aware $\BSTSGLTL$ yields a significant increase in detection
efficiency over $\BSGLsc$, even in the second data set where a transient single-detector disturbance and
transient signals are present together.
It achieves \mbox{$\hoscaledthr\approx0.08$} in both data sets and yields up to 35\% improvement in $\pDet$
for weak signals below this threshold.
This is also consistent with the expectations for a StackSlide search when taking into account
the additional mismatch accrued by semicoherent statistics.~\cite{wette2012:_sensitivity,wette2015:_metric}
\vspace{0.25\baselineskip}

Again, there are no losses with the simplified `loudest-only' detection statistic $\BSTSGLTLlo$ only using the
segment $\maxseg$ with \mbox{$\cohF^{\maxseg}=\max_{\segidx}\cohF^{\segidx}$} (not plotted separately).
\vspace{0.25\baselineskip}

However, a minor problem with $\BSTSGLTL$ is easy to overlook in \fref{fig:hsgct-simdata-detprobs-transient}.
Even at very high $h_0$, where other detection statistics eventually reach \mbox{$\pDet=1$}, it misses a few
signal injections, typically about \mbox{1\%--2\%}.
\vspace{0.25\baselineskip}

We found that all missed injections are related to the previously-mentioned peculiarities of data selection,
falling into segments 64 or 76, where one of the detectors contributes no or very little data.
\vspace{0.25\baselineskip}

Here, the single-segment hypotheses $\HypTLXk$ and $\HypTSk$ become indistinguishable,
and the normalization of $\BSTSGLTL$ with the given tuning values is such
that it will veto any strong outlier from these segments.
\vspace{0.25\baselineskip}

This issue is not a fundamental problem with our approach,
as the data selection for future semicoherent searches can easily be constrained
to avoid such anomalous segments,
making the search better suited for tCW detection
without risking efficiency for persistent CWs.
\vspace{0.25\baselineskip}

As a simple work-around for the given data selection, we can simply set
\mbox{$\oTLGXkargs{X}{\segidx=64,76}=0.0$}
while keeping all other $\oTLGXk$ at equal values.
This makes $\pDet(\BSTSGLTL)$ go to 1 for high $h_0$,
just as $\BSGLsc$ does,
and sacrifices only \mbox{1--2\%} of $\pDet$ at lower $h_0$,
and a similar small amount in the persistent-CW case
-- which could also be recovered by slightly changing the $\Ftho$ tuning.

\subsection{tCW signals with varying duration}
\label{sec:tests-TS-varduration}
The pragmatic signal model for transient CW-like signals from Eq.~\eref{eq:HypTSk} explicitly assumes a signal
lining up with a single segment, as tested in the previous section.
As such an alignment is not very likely in nature, it is interesting to test the robustness of the new
detection statistic against deviations from this assumption.
Hence, we have also tested injecting transient signals with random lengths $\Tinj$ and with random start times
uniformly drawn from $[\tstart,\tend-\Tinj]$.
\vspace{0.5\baselineskip}

\begin{figure}[t!]
  \includegraphics[width=\columnwidth,clip]{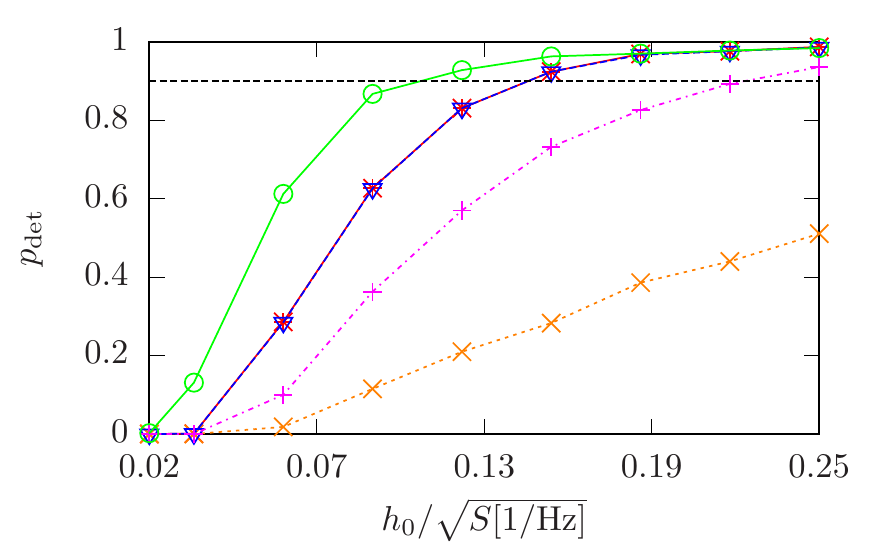} \\
  \includegraphics[width=\columnwidth,clip]{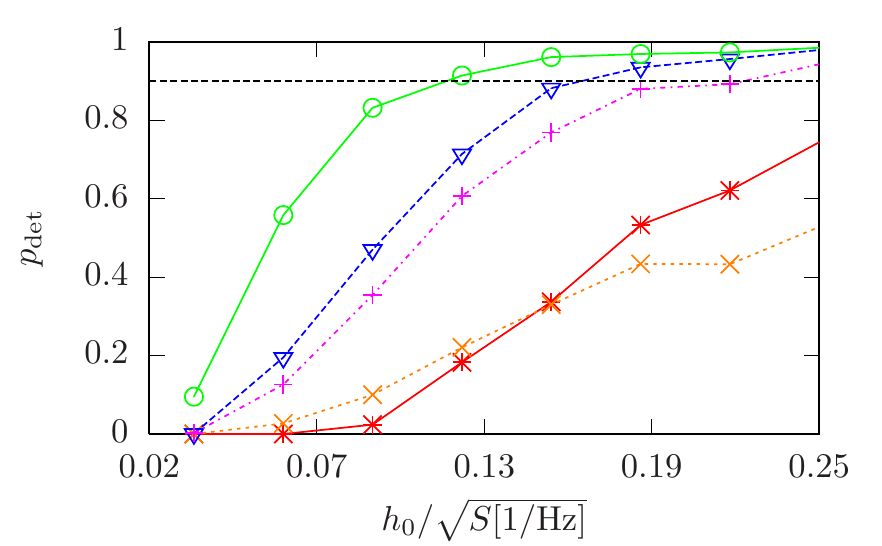}
  \caption{
    \label{fig:hsgct-simdata-detprobs-transient-rand}
    Detection efficiency $\pDet$ for
    transient tCW signals of random duration \mbox{$\Tinj\in[0.5,2.0]\times\Tseg$}
    and randomly distributed over the whole observation time.
    Here, the signal amplitude $h_0$ has been \emph{rescaled} for each injection to compensate for the
    varying $\Tinj$; see the text for details.
    Detection statistics and panels are the same as in
    Figs.~\ref{fig:hsgct-simdata-detprobs-persistent}--\ref{fig:hsgct-simdata-detprobs-transient}.
  }
\end{figure}

For better comparison with \fref{fig:hsgct-simdata-detprobs-transient}, the strength of these signals is
scaled according to \mbox{$h_0 \propto \sqrt{\Tseg/\Tinj}$},
so that the \emph{average} signal-to-noise ratios (\defineterm{SNRs})
for each nominal $h_0$ value are the same.
(But note that the per-injection SNRs are not fixed, as they still depend on the three randomized amplitude
parameters.)
\vspace{0.25\baselineskip}

As an example, we show the results for the Gaussian-noise data set, and for $\Tinj$ uniformly drawn from the
interval \mbox{$[0.5,2.0]\times\Tseg=[30,120]\hours$}, in \fref{fig:hsgct-simdata-detprobs-transient-rand}.
Here we find a general decrease in detection efficiency for all statistics considered, with $\scF$ and
$\BSGLsc$ only achieving \mbox{$\hoscaledthr\approx0.14$} instead of $\approx0.1$ in the previous test and
$\BSTSGLTL$ achieving \mbox{$\hoscaledthr\approx0.1$} instead of $\approx0.08$.
This is mostly due to the fact that now a significant fraction of injections falls completely or partially
within gaps of the simulated data set, or within parts with very low duty factor, thus decreasing the effective
SNR for any detection method.

The main finding here is that the transient-aware Bayes factor $\BSTSGLTL$ still improves over the other
statistics, and that this is still true even for the approximate `loudest-only' version.
This indicates that, though the initial assumption of a tCW signal that exactly matches the duration of a
single segment seemed quite strict and arbitrary, this simple approach is in fact useful for a wider range of
tCW durations.

We also observe that the permanence veto is not quite as strict in falsely vetoing these tCW signals as it was
for \mbox{$\Tinj=\Tseg$}, as some fraction of them that overlaps with more than one segment can now still
contribute to the reaveraged detection statistic.
Still, it is not competitive with any of the other tested detection statistics, which is no surprise, since it
was not constructed to accept transient signals in the first place.

\section{Conclusions}
\label{sec:conclusions}
\vspace{-0.5\baselineskip}
In this paper, we have considered $\F$-statistic-based semicoherent searches for CWs
on data that also contain transient instrumental disturbances
or signals that are transient, but also CW-like (quasimonochromatic).
We have generalized the Bayesian model-selection approach
of Refs.~\cite{prix09:_bstat,prix11:_transient,keitel2014:_linerobust}
with explicit models for transients lasting for a single segment in a semicoherent search,
including single-detector disturbances and multi-detector-coherent signals,
and demonstrated that the resulting statistic can be effective even for transients of different durations.

For this demonstration,
we injected simulated CW and tCW signals into several simulated data sets,
using realistic duty factors corresponding to the two LIGO detectors during their sixth science run.
We have shown that the new detection statistic $\BSTSGLTL$ is
\emph{more robust} than standard semicoherent statistics
towards both persistent and transient single-detector disturbances in the data,
while not hurting sensitivity to persistent CW signals,
and that it is \emph{more sensitive} towards transient signals.

We found that $\BSTSGLTL$ works best if the transient signals or disturbances indeed last for exactly a single
segment of the semicoherent search.
But it still yields significant improvement over standard methods for transient durations shorter or
longer than one segment length.

Though the injection studies showed a minor issue with $\BSTSGLTL$ dismissing a small number of strong tCW
signals, this was found to be related solely to a peculiarity of the data selection used in our tests.
It can be worked around by prior tuning,
while for future combined CW-tCW searches the issue can easily be avoided by a properly constrained data
selection.

Just as with the line-robust statistic $\BSGLsc$ of \PaperI,
the new detection statistic $\BSTSGLTL$ only requires quantities that are already computed
in any semicoherent $\scF$-statistic search:
namely the single-detector and single-segment $\cohF$-statistics.
Hence, computational cost is only increased by the arithmetic operations in computing $\BSTSGLTL$,
as given in Eq.~\eref{eq:lnBSGLTL}.
Our injection studies showed that this is usually dominated by a few terms,
and that good sensitivity can already be obtained by computing $\BSTSGLTL$ only for the most significant
candidates obtained from other statistics.
Hence, these results allow for a CW search with
increased robustness to transient disturbances
and increased sensitivity to transient signals
as a computationally cheap `add-on' to existing searches,
such as the Einstein@Home project~\cite{EatH}.

The present approach could be further generalized to allow for an explicit classification
of periodic data signatures into several classes
(persistent CW signals, persistent lines, transient disturbances, and `transient-CW' signals),
with transient lengths of any multiple of a segment length,
through the \textit{Bayesian Blocks} algorithm~\cite{scargle1998:_bayesblocks1,scargle2013:_bayesblocks2}.

Multi-detector transient disturbances cannot be safely distinguished 
from transient astrophysical signals
by considering the per-segment and per-detector $\F$-statistics only.
However, for widely separated detectors these are much rarer than single-detector disturbances,
and hence it should be possible to investigate any coincident transient candidates
with more detailed analysis of their frequency evolution and coherence with auxiliary channels.

As this work shows how to increase tCW sensitivity
by simple modifications to existing CW searches,
other more dedicated search methods might be more powerful,
but not without significant computational effort,
so that the present results will find practical applications regardless.
Still, sensitivity comparisons of tCW detection methods between this work,
the dedicated tCW detection statistic of Ref.~\cite{prix11:_transient}
and the more generic `long-transient' excess-power method 
of Ref.~\cite{thrane2015:_longstoch}
would be of great interest for the development of optimal overall search strategies.

Such a comparison will, however, require significant further work,
expertise from different sub-fields of GW data analysis,
and great care in ensuring a comprehensive and fair evaluation of each approach
under equal circumstances.
Hence, it would be an interesting opportunity for
a community mock-data challenge modelled after a recent study~\cite{scoX1MDC2015}
of methods to detect CW signals from the binary source Sco-X1,
which also compared CW-optimized and generic search methods.

\section*{Acknowledgments}
\vspace{-0.5\baselineskip}
I thank Reinhard Prix for valuable advice throughout this project and for feedback on the manuscript;
Berit Behnke, Maria Alessandra Papa, Ornella Piccinni, Irene Di Palma and Heinz-Bernd Eggenstein
for analyses of LIGO S5 and S6 data that inspired this study;
the LIGO Scientific Collaboration for providing those data sets;
Graham Woan, Ik Siong Heng and Eric Thrane for inspiring discussions about transient signals;
Sin{\'e}ad Walsh for advice on injection studies;
and HB Eggenstein for plotting help.
The injection studies on simulated data were carried out on the \mbox{ATLAS} cluster at AEI Hannover.
This paper has been assigned document number \mbox{\dcc{}}.

\appendix

\vspace{-\baselineskip}

\section{`cheat sheet':\\ pictorial representations of Bayes factors}
In this appendix, we provide a simple pictorial representation of the various detection statistics
(Bayes factors) derived from Bayesian hypothesis testing in Refs.~\cite{prix09:_bstat,keitel2014:_linerobust}
and in this paper.

Here we consider the simplest example case that still allows distinguishing all of our hypotheses $\HypGsc$,
$\HypSsc$, $\HypLsc$, $\HypTS$ and $\HypTL$:
this is the case of two detectors $X=1,2$ and two data segments $\segidx=1,2$.
We then represent with $\swsq$ any hypothesis that posits pure Gaussian noise
in the single-detector, single-segment data subset $(X,\segidx)$.
Alternatively, $\sbsq$ depicts any hypothesis that posits a quasimonochromatic signature,
be it a signal or a disturbance,
in $(X,\segidx)$.

We then build up sketches of the full semicoherent hypothesis by arranging the two detectors on the horizontal
axis and the two segments on the vertical axis.

This way, the signal-vs-Gaussian Bayes factor from Eq.~\eref{eq:OSG},
which is equivalent to the $\scF$-statistic~\cite{prix09:_bstat},
can be represented (up to proportionality, corresponding to global priors) as
\begin{equation}
 \label{eq:sketch_BSG}
 \hspace{-0.4cm}
 \BSGsc(\dVx)     \propto
 \eto{\scF(\dVx)} \propto
                   \frac{\prob{\HypSsc}{\dVx}}{\prob{\HypGsc}{\dVx}}
                  \propto
                   \frac{
                     \probBigl{\begin{smallmatrix} \sbsq & \sbsq \cr \sbsq & \sbsq \end{smallmatrix}}{\dVx}
                    }{
                     \probBigl{\begin{smallmatrix} \swsq & \swsq \cr \swsq & \swsq \end{smallmatrix}}{\dVx}
                    } \;.
\end{equation}

Using the same sketch notation, the pure line-veto statistic from \PaperI{} reads as
\begin{equation}
 \label{eq:sketch_BSL}
 \BSLsc(\dVx)    \propto \frac{
                     \probBigl{\begin{smallmatrix} \sbsq & \sbsq \cr \sbsq & \sbsq \end{smallmatrix}}{\dVx}
                  }{
                     \probBigl{\begin{smallmatrix} \sbsq & \swsq \cr \sbsq & \swsq \end{smallmatrix}}{\dVx}
                   + \probBigl{\begin{smallmatrix} \swsq & \sbsq \cr \swsq & \sbsq \end{smallmatrix}}{\dVx}
                  } \;,
\end{equation}
and the more general line-robust statistic, reproduced here in Eq.~\eref{eq:BSGL}, is
\begin{equation}
 \label{eq:sketch_BSGL}
 \hspace{-0.4cm}
 \BSGLsc(\dVx)   \propto \frac{
                     \probBigl{\begin{smallmatrix} \sbsq & \sbsq \cr \sbsq & \sbsq \end{smallmatrix}}{\dVx}
                  }{
                     \probBigl{\begin{smallmatrix} \swsq & \swsq \cr \swsq & \swsq \end{smallmatrix}}{\dVx}
                   + \probBigl{\begin{smallmatrix} \sbsq & \swsq \cr \sbsq & \swsq \end{smallmatrix}}{\dVx}
                   + \probBigl{\begin{smallmatrix} \swsq & \sbsq \cr \swsq & \sbsq \end{smallmatrix}}{\dVx}
                  } \;.
\end{equation}

In this paper, we have generalized the approach of \PaperI{} to yield
\onecolumngrid
\begin{enumerate}[(i)]
 \item a detection statistic that takes into account transient lines
       in any single-detector, single-segment subset as an additional noise component in the denominator,
       given in Eq.~\eref{eq:BSGLTL},
       and that we can sketch as
\begin{equation}
 \label{eq:sketch_BSGLTL}
 \BSGLTL(\dVx)   \propto \frac{
                      \probBigl{\begin{smallmatrix} \sbsq & \sbsq \cr \sbsq & \sbsq \end{smallmatrix}}{\dVx}
                   }{
                      \probBigl{\begin{smallmatrix} \swsq & \swsq \cr \swsq & \swsq \end{smallmatrix}}{\dVx}
                    + \probBigl{\begin{smallmatrix} \sbsq & \swsq \cr \sbsq & \swsq \end{smallmatrix}}{\dVx}
                    + \probBigl{\begin{smallmatrix} \swsq & \sbsq \cr \swsq & \sbsq \end{smallmatrix}}{\dVx}
                    + \probBigl{\begin{smallmatrix} \sbsq & \swsq \cr \swsq & \swsq \end{smallmatrix}}{\dVx}
                    + \probBigl{\begin{smallmatrix} \swsq & \sbsq \cr \swsq & \swsq \end{smallmatrix}}{\dVx}
                    + \probBigl{\begin{smallmatrix} \swsq & \swsq \cr \sbsq & \swsq \end{smallmatrix}}{\dVx}
                    + \probBigl{\begin{smallmatrix} \swsq & \swsq \cr \swsq & \sbsq \end{smallmatrix}}{\dVx}
                   } \;;
\end{equation}
 \item another detection statistic, given in Eq.~\eref{eq:BSTSGLTL},
       that also takes into account `transient-CW' signals in the numerator,
       which as a sketch looks like this:
\begin{align}
 \label{eq:sketch_BSTSGLTL}
 \BSTSGLTL(\dVx) \propto \frac{
                      \probBigl{\begin{smallmatrix} \sbsq & \sbsq \cr \sbsq & \sbsq \end{smallmatrix}}{\dVx}
                    + \probBigl{\begin{smallmatrix} \sbsq & \sbsq \cr \swsq & \swsq \end{smallmatrix}}{\dVx}
                    + \probBigl{\begin{smallmatrix} \swsq & \swsq \cr \sbsq & \sbsq \end{smallmatrix}}{\dVx}
                   }{
                      \probBigl{\begin{smallmatrix} \swsq & \swsq \cr \swsq & \swsq \end{smallmatrix}}{\dVx}
                    + \probBigl{\begin{smallmatrix} \sbsq & \swsq \cr \sbsq & \swsq \end{smallmatrix}}{\dVx}
                    + \probBigl{\begin{smallmatrix} \swsq & \sbsq \cr \swsq & \sbsq \end{smallmatrix}}{\dVx}
                    + \probBigl{\begin{smallmatrix} \sbsq & \swsq \cr \swsq & \swsq \end{smallmatrix}}{\dVx}
                    + \probBigl{\begin{smallmatrix} \swsq & \sbsq \cr \swsq & \swsq \end{smallmatrix}}{\dVx}
                    + \probBigl{\begin{smallmatrix} \swsq & \swsq \cr \sbsq & \swsq \end{smallmatrix}}{\dVx}
                    + \probBigl{\begin{smallmatrix} \swsq & \swsq \cr \swsq & \sbsq \end{smallmatrix}}{\dVx}
                   } \;;
\end{align}
 \item a pure transient-CW detection statistic, as given in Eq.~\eref{eq:BTSGLTL}, and with the following sketch form:
\begin{align}
 \label{eq:sketch_BTSGLTL}
 \BTSGLTL(\dVx) \propto \frac{
                    \probBigl{\begin{smallmatrix} \sbsq & \sbsq \cr \swsq & \swsq \end{smallmatrix}}{\dVx}
                    + \probBigl{\begin{smallmatrix} \swsq & \swsq \cr \sbsq & \sbsq \end{smallmatrix}}{\dVx}
                   }{
                      \probBigl{\begin{smallmatrix} \swsq & \swsq \cr \swsq & \swsq \end{smallmatrix}}{\dVx}
                    + \probBigl{\begin{smallmatrix} \sbsq & \swsq \cr \sbsq & \swsq \end{smallmatrix}}{\dVx}
                    + \probBigl{\begin{smallmatrix} \swsq & \sbsq \cr \swsq & \sbsq \end{smallmatrix}}{\dVx}
                    + \probBigl{\begin{smallmatrix} \sbsq & \swsq \cr \swsq & \swsq \end{smallmatrix}}{\dVx}
                    + \probBigl{\begin{smallmatrix} \swsq & \sbsq \cr \swsq & \swsq \end{smallmatrix}}{\dVx}
                    + \probBigl{\begin{smallmatrix} \swsq & \swsq \cr \sbsq & \swsq \end{smallmatrix}}{\dVx}
                    + \probBigl{\begin{smallmatrix} \swsq & \swsq \cr \swsq & \sbsq \end{smallmatrix}}{\dVx}
                   } \;.
\end{align}
\end{enumerate}

\twocolumngrid

\bibliography{../../biblio.bib}

\end{document}